\documentclass[twocolumn,preprintnumbers,amsmath,amssymb,nobibnotes,nofootinbib,superscriptaddress]{revtex4}
\usepackage{graphicx}% Include figure files
\usepackage{dcolumn}% Align table columns on decimal point
\usepackage[usenames,dvipsnames]{color}
\RequirePackage[colorlinks=true
,urlcolor=blue
,anchorcolor=blue
,citecolor=blue
,filecolor=blue
,linkcolor=blue
,menucolor=blue
,pagecolor=blue
,linktocpage=true
,pdfproducer=medialab
]{hyperref}
\usepackage{amsmath}
\usepackage{mathrsfs}
\usepackage{slashed}
\usepackage{cancel}

\newcommand{\LL}{\mathscr{L}}
\newcommand{\cG}{\mathscr{G}}

\def\cD{{\cal D}}
\def\cF{{\cal F}}

\def\cG{{\cal G}}

\def\cO{{\cal O}}
\def\cOL{{\cal O}_L}
\def\cOR{{\cal O}_R}
\def\cOLR{{\cal O}_{L(R)}}
\def\cOLt{\widetilde{{\cal O}}_L}
\def\cORt{\widetilde{{\cal O}}_R}
\def\cOLRt{\widetilde{{\cal O}}_{L(R)}}
\def\cP{{\cal P}}
\def\cS{{\cal S}}

\def\Tr{{\rm Tr}}

\def\be{\begin{equation}}
\def\ee{\end{equation}}
\def\beq{\begin{equation}}
\def\eeq{\end{equation}}
\def\bc{\begin{center}}
\def\ec{\end{center}}
\def\bea{\begin{eqnarray}}
\def\eea{\end{eqnarray}}

\def\nt{\noindent}

\newcommand{\derp}{\partial}

\newcommand{\LT}{\mathbf{L}}
\newcommand{\RT}{\mathbf{R}}
\newcommand{\UY}{\mathbf{U_Y}}
\newcommand{\UH}{\mathbf{U}}
\newcommand{\UHL}{\mathbf{U}_L}
\newcommand{\UHR}{\mathbf{U}_R}
\newcommand{\UHLR}{\mathbf{U}_{L(R)}}

\newcommand{\TL}{\mathbf{T}}
\newcommand{\VL}{\mathbf{V}}
\newcommand{\DL}{D}

\newcommand{\DLL}{\mathcal{D}}

\newcommand{\VLLmuu}{\mathbf{V}^\mu_L}
\newcommand{\VLLmud}{\mathbf{V}_{\mu,\,L}}

\newcommand{\VLRmuu}{\mathbf{V}^\mu_R}

\newcommand{\VLLmuut}{\widetilde{\mathbf{V}}^\mu_L}
\newcommand{\VLLmudt}{\widetilde{\mathbf{V}}_{\mu,\,L}}
\newcommand{\VLLnuut}{\widetilde{\mathbf{V}}^\nu_L}
\newcommand{\VLLnudt}{\widetilde{\mathbf{V}}_{\nu,\,L}}
\newcommand{\VLRmuut}{\widetilde{\mathbf{V}}^\mu_R}
\newcommand{\VLRmudt}{\widetilde{\mathbf{V}}_{\mu,\,R}}
\newcommand{\VLRnuut}{\widetilde{\mathbf{V}}^\nu_R}
\newcommand{\VLRnudt}{\widetilde{\mathbf{V}}_{\nu,\,R}}
\newcommand{\VLLrhout}{\widetilde{\mathbf{V}}^\rho_L}

\newcommand{\VLRrhout}{\widetilde{\mathbf{V}}^\rho_R}

\newcommand{\BBu}{B^{\mu\nu}}
\newcommand{\BBd}{B_{\mu\nu}}

\newcommand{\WWLut}{\widetilde{W}^{\mu\nu}_L}
\newcommand{\WWRut}{\widetilde{W}^{\mu\nu}_R}

\newcommand{\WWWRut}{\widetilde{W}^{\rho\sigma}_R}

\newcommand{\TLchi}{\mathbf{T}_\chi}
\newcommand{\VLchimuu}{\mathbf{V}^\mu_\chi}
\newcommand{\VLchimud}{\mathbf{V}_{\mu,\,\chi}}
\newcommand{\VLchinuu}{\mathbf{V}^\nu_\chi}
\newcommand{\VLchinud}{\mathbf{V}_{\nu,\,\chi}}
\newcommand{\VLchirhou}{\mathbf{V}^\rho_\chi}

\newcommand{\WWchiu}{W^{\mu\nu}_\chi}

\newcommand{\WWWchiu}{W^{\rho\sigma}_\chi}

\newcommand{\TLLR}{\mathbf{T}_{L(R)}}
\newcommand{\VLLRmuu}{\mathbf{V}^\mu_{L(R)}}
\newcommand{\VLLRnuu}{\mathbf{V}^\nu_{L(R)}}
\newcommand{\TLLt}{\widetilde{\mathbf{T}}_{L}}
\newcommand{\TLRt}{\widetilde{\mathbf{T}}_{R}}
\newcommand{\TLLRt}{\widetilde{\mathbf{T}}_{L(R)}}
\newcommand{\VLLRmuut}{\widetilde{\mathbf{V}}^\mu_{L(R)}}
\newcommand{\WWLRu}{W^{\mu\nu}_{L(R)}}
\newcommand{\WWLRut}{\widetilde{W}^{\mu\nu}_{L(R)}}

\newcommand{\eps}{\epsilon_{\mu\nu\rho\sigma}}
\newcommand{\fL}{f_L}
\newcommand{\fR}{f_R}
\newcommand{\fLR}{f_{L(R)}}
\newcommand{\gL}{g_L}
\newcommand{\gR}{g_R}
\newcommand{\gLR}{g_{L(R)}}
\newcommand{\gchi}{g_\chi}

\newcommand{\brown}[1]{\color[rgb]{0.8,0.1,0.} #1 \color{black}}

\begin{document}
%
%%%
%%%%%%%%%%%%%%%%%%% Title Page%%%%%%%%%%%%%%%%%%%%%%%%%
%%%
%

\title{CP violation from spin--1 resonances in a left--right dynamical Higgs context}

\author{Kunming Ruan}
\email{runkunming@itp.ac.cn}
\affiliation{State Key Laboratory of Theoretical Physics and Kavli Institute for Theoretical Physics China (KITPC)\\
Institute of Theoretical Physics, Chinese Academy of Sciences, Beijing 100190, P. R. China}

\author{Jing Shu}
\email{jshu@itp.ac.cn}
\affiliation{State Key Laboratory of Theoretical Physics and Kavli Institute for Theoretical Physics China (KITPC)\\
Institute of Theoretical Physics, Chinese Academy of Sciences, Beijing 100190, P. R. China}

\author{Juan Yepes}
\email{juyepes@itp.ac.cn}
\affiliation{State Key Laboratory of Theoretical Physics and Kavli Institute for Theoretical Physics China (KITPC)\\
Institute of Theoretical Physics, Chinese Academy of Sciences, Beijing 100190, P. R. China}

\begin{abstract}

New physics field content in the nature, more specifically, from spin--1 resonances sourced by the extension of the SM local gauge symmetry to the larger local group $SU(2)_L\otimes SU(2)_R\otimes U(1)_{B-L}$, may induce CP--violation signalling NP effects from higher energy regimes. In this work we completely list and study all the CP--violating operators up to the $p^4$--order in the Lagrangian expansion, for a non--linear left--right electroweak chiral context and coupled to a light dynamical Higgs. Heavy right handed fields can be integrated out from the physical spectrum, inducing thus a physical impact in the effective gauge couplings, fermionic electric dipole moment, and CP-violation in the
decay $h\rightarrow ZZ^*\rightarrow 4\,l$ that are briefly analysed. The final relevant set of effective operators have also been identified at low energies.

\end{abstract}
\maketitle

%
%%%%%%%%%%%%%%%%%%%%%%%%%   1.  Introduction       %%%%%%%%%%%%%%%%%%%%%%%%
%
\section{Introduction}

\nt The Standard Model (SM) has been finally established as a coherent and consistent picture of electroweak symmetry breaking (EWSB) after the LHC experimental confirmation of a new scalar resonance~\cite{:2012gk,:2012gu} in our nature, that resembles the long ago proposed Higgs boson particle~\cite{Englert:1964et,Higgs:1964ia,Higgs:1964pj}. Nevertheless, new physics (NP) effects are still awaiting to be detected as it is demanded by the long--standing hierarchy problem in particle physics. Among those possible NP manifestations, CP--violating low energy effects could prove the existence of higher energy regimes, reachable and detectable at the LHC and future facilities. Indeed, the electroweak interactions taking place in our nature do not conserve completely the product of the charge conjugation and parity symmetries of particle physics. Moreover, the observed matter--antimatter asymmetry of our universe compel us to consider new sources of CP--violation, that are signalled as well by the extreme fine-tuning entailed by the strong CP problem. 

Phenomenological analysis pursuing effective signals have been performed~\cite{Dell'Aquila:1985ve,Dell'Aquila:1985vc,Dell'Aquila:1985vb,
Soni:1993jc,Chang:1993jy,Arens:1994wd,Choi:2002jk,Buszello:2002uu,
Godbole:2007cn,Cao:2009ah,Gao:2010qx,DeRujula:2010ys,Plehn:2001nj,
Buszello:2006hf,Hankele:2006ma,Klamke:2007cu,Englert:2012ct,
Odagiri:2002nd,DelDuca:2006hk,Andersen:2010zx,Englert:2012xt,
Djouadi:2013yb,Dolan:2014upa,Christensen:2010pf,Desai:2011yj,
Ellis:2012xd,Godbole:2013saa,Delaunay:2013npa,Voloshin:2012tv,
Korchin:2013ifa,Bishara:2013vya,Chen:2014ona,Freitas:2012kw,
Djouadi:2013qya,Belusca-Maito:2014dpa,Shu:2013uua,Cheung:2013kla,Dwivedi:2015nta} in order to search for anomalous CP--odd fermion--Higgs and gauge--Higgs
couplings. Complementary studies from the theoretical side are very relevant to establish and analyse the complete set of independent
CP--violating bosonic operators, as they may shed a direct light
on the nature of EWSB and pinpoint NP effects from higher energy regimes.

Motivated by these tantalizing and challenging prospects, this work copes with the possibility for detecting non--zero CP--violating signals arising out from some emerging new physics field content in the nature, more specifically, from spin--1 resonances brought into the game via the extension of the SM local gauge symmetry $\cG_{SM}=SU(2)_L\otimes U(1)_Y$ up to the larger local group $\cG=SU(2)_L\otimes SU(2)_R\otimes U(1)_{B-L}$ (see~\cite{LRSM1,LRSM2} for left-right symmetric models literature). Such extended gauge field sector is tackled here through a systematic and model-independent EFT approach. The basic strategy is to employ a non--linear $\sigma$--model to account for the strong dynamics giving rise to the GB, that is the $W^\pm_L$ and $Z_L$ longitudinal components that leads to introduce the Goldstone scale $\fL$, together with the corresponding GB from the extended local group, i.e. the additional $W^\pm_R$ and $Z_R$ longitudinal degrees of freedom and the associated Goldstone scale $\fR$. Finally, this non--linear $\sigma$--model effective Lagrangian is coupled a posteriori to a scalar singlet $h$ in a general way through powers of $h/\fL$~\cite{Georgi:1984af}, being the scale suppression dictated by $\fL$, as it is the scale where $h$ is generated as a GB prior to the extension of the SM local group $\cG_{SM}$ to the larger one $\cG$.  

In this work we analysed the physical picture of spin--1 resonances dictated by the larger local gauge group $\cG$, with an underlying strong interacting scenario coupled to a light Higgs particle, via the non--linear EFT construction of the complete tower of pure gauge and gauge-$h$ operators up to the $p^4$--order in the Lagrangian expansion and restricted only to the CP non--conserving bosonic sector. The corresponding CP--conserving counterpart was recently analysed in~\cite{Yepes:2015zoa}. The work in here enlarges and completes the operator basis previously considered in~\cite{Zhang:2007xy,Wang:2008nk} (for the CP--breaking sector) in the context of left--right symmetric EW chiral models, completing and generalizing also the work done in~\cite{Appelquist:1980vg,Longhitano:1980iz,Longhitano:1980tm,
Feruglio:1992wf,Appelquist:1993ka} (for the CP--violating sector), and it extends as well~\cite{Gavela:2014vra} to the case of a larger local gauge symmetry $\cG$ in the context of non--linear EW interactions coupled to a light Higgs particle. The theoretical framework undertaken in here may be considered as well as a generic UV completion of the low energy non--linear approaches of~\cite{Appelquist:1980vg,Longhitano:1980iz,Longhitano:1980tm,
Feruglio:1992wf,Appelquist:1993ka}  and~\cite{Gavela:2014vra}.

%%%%%%%%%%%%%%%%%%%%%%  2. Theoretical framework  %%%%%%%%%%%%%%%%%%

\section{Theoretical framework}
\label{Theoretical-framework}

\nt The transformation properties of the longitudinal  degrees of freedom of the electroweak gauge bosons will be parametrized at low-energies as it is customary via the dimensionless unitary matrix $\UH(x)$, more specifically through $\UHL(x)$ and $\UHR(x)$ for the symmetry group $SU(2)_L\otimes SU(2)_R$, and defined as
\beq
\UHLR\,(x)=e^{i\,\tau_a\,\pi^a_{L(R)}(x)/\fLR}\, , 
\label{Goldstone-matrices}
\eeq
\nt with $\tau^a$ Pauli matrices and $\pi^a_{L(R)}(x)$ the corresponding Goldstone bosons fields suppressed by their associated non--linear sigma model scale $\fLR$, where the scale $f_R$ comes from the additional Goldstone boson dynamics introduced by $SU(2)_R$ group. It is customary to introduce the corresponding covariant derivative objects for both of the Goldstone matrices $\UHLR(x)$ as
\be
\begin{aligned}
&\DL^\mu \UHLR \equiv \\
&\derp^\mu \UHLR \, + \,\frac{i}{2}\,\gLR\,W^{\mu,a}_{L(R)}\,\tau^a\,\UHLR - \frac{i}{2}\,g'\,B^\mu\, \UHLR\,\tau^3 \,.
\end{aligned}
\label{Covariant-derivatives}
\ee

\nt with the $SU(2)_L$, $SU(2)_R$ and $U(1)_{B-L}$ gauge fields denoted by $W^{a\mu}_L$, $W^{a\mu}_R$ and $B^\mu$ correspondingly, and the associated gauge couplings $\gL$, $\gR$ and $g'$ respectively. Additionally, it is straightforward to introduce in the framework the adjoints $SU(2)_{L(R)}$--covariant vectorial $\VLLRmuu$ and 
the covariant scalar $\TLLR$ objects as
\be
\VLLRmuu \equiv \left(\DL^\mu \UHLR\right)\,\UHLR^\dagger\,, \quad
\TLLR \equiv \UHLR\,\tau_3\,\UHLR^\dagger\,,
\label{EFT-building-blocks}
\ee
\nt covariantly transforming all them under local 
transformations of the larger group $\cG$. 

Notice that the local gauge invariance of the theory allows to build operators made out of traces depending either on products of purely left-handed or right-handed covariant objects. As soon as the operators mixing the left and right-handed structures are considered, new covariant objects emerge to fully guarantee their construction. In fact, considering for instance the simple trace $\Tr\left(\cOL^i\,\cOR^j\right)$ mixing the left and right--handed covariant objects $\cOL^i$ and $\cOR^i$ respectively (with $i$ labelling either a scalar, vector, or tensor object), the gauge invariance no longer holds. Proper insertions of the Goldstone matrices $\UHL$ and $\UHR$ will make it invariant through $\Tr\left(\cOL^i\,\cOR^j\right) \rightarrow\Tr\left(\UHL^\dagger\,\cOL^i\,\UHL\,\UHR^\dagger\,\cOR^j\,\UHR\right)$, motivating thus the introduction of the following objects
\be
\cOLt^i \equiv \UHL^\dagger\,\cOL^i\,\UHL\,, \qquad\qquad
\cORt^i \equiv \UHR^\dagger\,\cOR^i\,\UHR\,,
\label{U-transforming-objects}
\ee

\nt that are required hereafter for the construction of operators made out of mixed $SU(2)_L$ and $SU(2)_R$ covariant structures. Notice that under the local $\cG$--transformations 
\be
\LT(x)\equiv e^{\frac{i}{2}\tau^a\alpha^a_L(x)},\, 
\RT(x)\equiv e^{\frac{i}{2}\tau^a\alpha^a_R(x)},\, 
\UY(x)\equiv e^{\frac{i}{2}\tau^3\alpha^0(x)}
\label{Local-transformations}
\ee
\nt with $\alpha^a_{L,R}(x)$ and $\alpha^0(x)$ space-time dependent variables parametrizing the local rotations, the new defined objects in Eq.~\eqref{U-transforming-objects} are transforming as
\be
\cOLt^i \rightarrow \mathbf{U}_Y\,\cOLt^i\,\mathbf{U}^\dagger_Y\,, \qquad\qquad
\cORt^i \rightarrow \mathbf{U}_Y\,\cORt^i\,\mathbf{U}^\dagger_Y
\label{Utilde-transformation-properties}
\ee
\nt The corresponding definitions in~\eqref{U-transforming-objects} for the covariant vectorial $\VLLRmuu$ and the scalar $\TLLR$ objects in~\eqref{EFT-building-blocks} are
\be
\begin{aligned}
\VLLRmuut &\equiv \UHLR^\dagger\,\VLLRmuu\,\UHLR = -\left(\DL^\mu \UHLR \right)^\dagger\UHLR\,,
\end{aligned}
\label{Vtilde}
\ee
\be
\TLLRt \equiv \UHLR^\dagger\,\TLLR\,\UHLR = \tau_3\,,
\label{Ttilde}
\ee
\nt where the unitary property of the Goldstone matrices $\UHLR$  has been employed in addition. Similar definitions for the strength gauge fields $\WWLRu$ are straightforward
\be
\WWLRut \equiv \UHLR^\dagger\,\WWLRu\,\UHLR\,.
\label{Wtilde}
\ee
\nt It is possible to infer therefore the mandatory introduction of the covariant objects $\cOLRt^i$ in order to construct any possible operator mixing the left and right-handed covariant quantities $\cOLR^i$. As it was realized in~\cite{Yepes:2015zoa}, operators made out of products of purely left or right-handed covariant quantities can be constructed out also via the covariant objects $\cOLRt^i$ defined in Eq.~\eqref{U-transforming-objects}. Henceforth the set  $\{\VLLRmuut,\,\TLLRt\}$ (together with Eq.~\eqref{Wtilde}) will make up the building blocks for the construction of the effective EW non--linear left-right CP--violating approach undertaken in this work, whose corresponding CP--conserving counterpart was already explored in Ref.~\cite{Yepes:2015zoa}. Construction that will be enlarged after accounting for all the possible gauge-Higgs couplings arising out in this scenario via the generic polynomial light Higgs function $\cF(h)$~\cite{Alonso:2012px}, singlet under $\cG$, and defined through the generic expansion 
\beq
\cF_i(h)\equiv1+2\,{a}_i\,\frac{h}{\fL}+{b}_i\,\frac{h^2}{\fL^2}+\ldots\,,
\label{F}
\eeq
\nt with dots standing for higher powers in $h/\fL$~\cite{Georgi:1984af}, not considered below. The scale suppression for each $h$--insertion is  dictated by $\fL$, as it is the scale where $h$ is generated as a GB prior to the extension of the SM local group $\cG_{SM}$ to the larger one $\cG$. Gauge--$h$ interactions will arise by letting the non--linear operators to be coupled either directly to $\cF(h)$ or through its derivative couplings, e.g. via $\partial^\mu \cF(h)$, $\left(\partial^\mu \cF(h)\right)^2$, $\partial^\mu \cF(h)\,\partial_\mu \cF(h)$ and $\partial^\mu\partial_\mu \cF(h)$. Thus the building blocks set gets complemented by accounting for all the possible interactions from $\cF(h)$  and its corresponding derivative couplings, under the assumption of a CP--even behaviour for $h$.

The local gauge symmetry $\cG$ has been demanded throughout the non--linear effective approach considered up to now. Another relevant symmetry emerges once the initial SM local group $\cG_{SM}$ is enlarged to $\cG$ and related with the exchange of the left and right--handed components of the group $SU(2)_L\otimes SU(2)_R$\,: the discrete parity symmetry $P_{LR}$, useful for protecting the $Zb\bar b$--coupling from large corrections in the context of composite Higgs models~\cite{Agashe:2006at}, and helpful in here for bringing more effective operators in the scenario as it will be realized when listing the operators afterwards.

\nt The CP-violating set-up described in the next section follows the dynamical Higgs scenario in~\cite{Alonso:2012px,Brivio:2013pma,Gavela:2014vra,Alonso:2012pz} (see also~\cite{Buchalla:2013rka,Buchalla:2013eza}, and for a short summary~\cite{Brivio:2015hua}), as well as the left-right bosonic\footnote{See~\cite{Cvetic:1988ey,Alonso:2012jc,Alonso:2012pz,
Buchalla:2012qq,Buchalla:2013rka} for non--linear analysis including fermions.} CP-conserving picture in~\cite{Yepes:2015zoa}.

%%%%%%%%%%%%%%%%%%%  3. The effective Lagrangian  %%%%%%%%%%%%%%%%%%

\section{The effective Lagrangian}
\label{EffectiveLagrangian}

\nt The effective CP--violating NP contributions from the strong dynamics assumed in here will lead to non-zero departures with respect to the SM Lagrangian $\LL_0$ and will be encoded in the Lagrangian $\LL_\text{chiral}$ through
\beq
\LL_\text{chiral} = \LL_0\,+\,\LL_{0,LR}
\,+\,\Delta \LL_{\text{\cancel{CP}}}\,+\,\Delta\LL_{\text{\cancel{CP}},LR}\,.
\label{Lchiral}
\eeq 

\nt Concerning only the bosonic interacting sector, the first piece in $\LL_\text{chiral}$ reads as
\be
\begin{aligned}
&\LL_0 = -\dfrac{1}{4}\,\BBd\,\BBu-\dfrac{1}{4}\,W^a_{\mu\nu,\,L}\,W^{\mu\nu,\,a}_L-
\dfrac{1}{4}\,G^a_{\mu\nu}\,G^{\mu\nu,\,a}\,+\\[2mm]
&+\frac{1}{2} (\derp_\mu h)(\derp^\mu h) - V (h)-\dfrac{\fL^2}{4}\Tr\Big(\VLLmuu\VLLmud\Big)\left(1+\frac{h}{\fL}\right)^2+\\
&-\frac{g_s^2}{32\,\pi^2}\,\theta_s\,\epsilon^{\mu\nu\rho\sigma}\,G^a_{\mu\nu}\, G^a_{\rho\sigma}\,,
\end{aligned}
\label{LLO}
\ee

\nt providing the SM strength gauge kinetic terms canonically normalized in the first line, whereas $h$--kinetic terms and the effective scalar potential $V(h)$ triggering the EWSB from the first two terms at the second line, plus the $W^\pm_L$ and $Z_L$ masses\footnote{As long as the corresponding right handed and mixed left--right handed terms are regarded, as in Ref.~\cite{Yepes:2015zoa}, mass mixing terms either for the neutral or charged sector are induced~\cite{Shu:2015cxm}.} and their couplings to the scalar $h$ together with the GB--fermion kinetic terms from the remaining piece in the second line of Eq.~\eqref{LLO}. Finally, the last term in Eq.~(\ref{LLO}) corresponds to the well-known total derivative CP-odd gluonic coupling. Notice that the $SU(2)_L$--kinetic term and the custodial conserving $p^2$--operator $\Tr\left(\VL^\mu\,\VL_\mu\right)$ have been properly labelled in order to keep clear the notation according to the assumed local symmetry group $\cG$. GB--kinetic terms are already canonically normalized from the scale factor of $\Tr\left(\VLLmuu\,\VLLmud\right)$, in agreement with $\UHL$--definition of Eq.~\eqref{Goldstone-matrices}.

Non--zero NP departures with  respect to those from $\LL_0$ will play a role into the game once the symmetric counterpart sourced by the corresponding local $SU(2)_R$--extension is called in, being encoded through the remaining pieces of $\LL_\text{chiral}$ Eq.~\eqref{Lchiral}, focused only on the CP--violating operators, and defined by

\begin{itemize}

\item $\LL_{0,LR}$: accounting for all of the possible $p^2$--operators mixing the left and right handed-covariant objects, 

\item $\Delta\LL_{\text{\cancel{CP}}}$: encoding all those operators up to the $p^4$--contributions made out of purely left or right handed covariant objects, and 

\item $\Delta\LL_{\text{\cancel{CP}},LR}$: parametrising any possible mixing interacting terms between $SU(2)_L$ and $SU(2)_R$--covariant objects up to $p^4$--operators permitted by the underlying left-right symmetry. 

\end{itemize}

\nt It is worth to comment that at the LO $p^2$--order, the corresponding CP--violating $SU(2)_R$--strength gauge kinetic term would be encoded in the Lagrangian $\LL_{0,R}$, turns out to be proportional to a cuadridivergence, being thus disregarded from the Lagrangian $\LL_\text{chiral}$ in Eq.~\eqref{Lchiral}. As soon as the light Higgs couplings are switched on through $\cF(h)$ in~\eqref{F}, such $SU(2)_R$ kinetic term has to be retained in the effective approach as it will be done later soon.

%%%%%%%%%%%%%%%%%%%%%%%%%  3.1 \LL_{0,LR}  %%%%%%%%%%%%%%%%%%%%%%%%%%

\boldmath
\subsection{$SU(2)_L-SU(2)_R$ $p^2$--interplay: $\LL_{0,LR}$}
\unboldmath

\nt The LO $p^2$--Lagrangian for the $SU(2)_R$--extension of the framework brings operators made of mixed products of left and right--handed objects via the covariant objects $\cOLRt$ defined in Eq.~\eqref{U-transforming-objects}, specifically from $\VLLRmuut$, $\TLLRt$ and $\WWLRut$ defined in Eqs.~\eqref{Vtilde}, \eqref{Ttilde} and \eqref{Wtilde} respectively. In fact, the $p^2$--interplaying Lagrangian for such contributions is parametrized by 
\be
\LL_{0,LR}= -\dfrac{1}{2}\,\eps\Tr\left(\WWLut\,\WWWRut\right)\,.
\label{LLO-Left-Right}
\ee

\nt Contrary to the case of either the left or right handed strength gauge kinetic terms, the latter operator has to be maintained in the effective Lagrangian, as it can not be traded by a cuadridivergence due to the mixed fields involved in Eq.~\eqref{LLO-Left-Right}.

At higher orders in the momentum expansion more interactions are sourced by the local symmetry group $\cG$, part of them accounted by the third piece of $\LL_\text{chiral}$ in Eq.~\eqref{Lchiral}, i.e.  $\Delta \LL_{\text{\cancel{CP}}}$, and described in the following.

%%%%%%%%%%%%%%%% 3.2 \Delta\LL_{\cancel{\text{CP}}}  %%%%%%%%%%%%%%%%

\boldmath
\subsection{$\cG$--extension of $\LL_0\,+\,\LL_{0,LR}$: $\Delta\LL_{\cancel{\text{CP}}}$}
\unboldmath

\nt  $\Delta\LL_{\cancel{\text{CP}}}$ describes deviations from the LO Lagrangian $\LL_0\,+\,\LL_{0,LR}$, encoding all the possible CP--violating gauge--$h$ interactions up to $p^4$--operators, and are split in here as
\be
\Delta \LL_{\text{\cancel{CP}}}=\Delta \LL_{\text{\cancel{CP}},L}+\Delta \LL_{\text{\cancel{CP}},R}
\label{DeltaL-CP-odd}
\ee
\nt with the suffix $L(R)$ labelling all those operators built up via the $SU(2)_{L(R)}$ building blocks provided in Section~\ref{Theoretical-framework}. In the context of purely EW chiral effective theories coupled to a light Higgs, the first contribution to $\Delta \LL_{\cancel{\text{CP}}}$, i.e. $\Delta \LL_{\cancel{\text{CP}},L}$, has already been provided in Refs.~\cite{Gavela:2014vra}, whereas part of $\Delta \LL_{\cancel{\text{CP}},L}$ and $\Delta \LL_{\cancel{\text{CP}},R}$ were already analysed for the left--right symmetric frameworks in Refs.~\cite{Zhang:2007xy,Wang:2008nk}. Both of the contributions in Eq.~\eqref{DeltaL-CP-odd} can be correspondingly written down as 
\beq
\hspace*{-0.6cm}
\begin{aligned}
&\Delta \LL_{\text{\cancel{CP}},L}=\\
&c_G\,\cS_{G}(h) + c_B\,\cS_{B}(h) + \sum_{i=\{W,2D\}}c_{i,L}\,\cS_{i,L}(h)+\sum_{i=1}^{16}c_{i,L}\,\cS_{i,L}(h)
\label{DeltaL-CP-odd-L}
\end{aligned}
\eeq
\nt and
\be
\Delta \LL_{\text{\cancel{CP}},R}=\sum_{i=\{W,2D\}}c_{i,R}\,\cS_{i,R}(h)\,\,+\,\,\sum_{i=1}^{16}\,c_{i,R}\,\cS_{i,R}(h),
\label{DeltaL-CP-odd-R}
\ee
\nt where the coefficients $c_{B}$, $c_{G}$ and  $c_{i,\chi}$, (with $\chi$ standing for $\chi=L,R$)  are model--dependent constant coefficients, whilst the first three terms of $\Delta \LL_{\text{\cancel{CP}},L}$ in Eq.~\eqref{DeltaL-CP-odd-L} and the first term in Eq.~\eqref{DeltaL-CP-odd-R} can be jointly written as
\be
\begin{aligned}
\cS_{G}(h)&= -\frac{1}{4}\,g^2_s\,\epsilon^{\mu\nu\rho\sigma}\,G^{a\mu\nu}\, G^a_{\rho\sigma}\,\cF_{G}(h)\,,\\
\cS_{B}(h)&= -\frac{1}{4}\,g'^2\,\epsilon^{\mu\nu\rho\sigma}\,B^{\mu\nu}\, B_{\rho\sigma}\,\cF_{B}(h)\,,\\ 
\cS_{W,\chi}(h)&= -\frac{1}{4}\,\gchi^2\,\eps\,\Tr\Big(\WWchiu\,\WWWchiu\Big)\cF_{W,\chi}(h)\,,\\ 
\cS_{2D,\chi}(h)&= i\,\frac{\fL^2}{4}\,\Tr\Big(\TL\,\cD_\mu\VLchimuu\Big)\,\cF_{2D,\chi}(h)\,,
\end{aligned}
\label{G-2D}
\ee 

\nt with suffix $\chi$ labelling again as $\chi=L,R$, and the generic $\cF_i(h)$--function of the scalar singlet $h$ is defined for all the operators following Eq.~\eqref{F}. The Higgs--independent terms are physically irrelevant for operators $\cS_{B}(h)$, $\cS_{W,\chi}(h)$ and $\cS_{2D}(h)$ as the first two operators can be written in terms of a cuadridivergence and the latter one turns out to be vanishing after integration by parts. The covariant derivative $\DLL_\mu$ of a field transforming in the adjoint representation of $SU(2)_L$ is defined as
\be
\DLL^\mu \VLchinuu \equiv \partial^\mu \VLchinuu\,+\,i\,\gchi\left[ W^\mu_\chi, \VLchinuu \right]\,,\quad \chi=L,R\,.
\label{DV-covariant-derivative}
\ee

\nt The complete linearly independent set of 16 CP--violating pure gauge and gauge--$h$ non-linear $\cG$--invariant operators and up to the $p^4$-order in the effective Lagrangian expansion, are encoded by $\cS_{i,\,L}(h)$ (fourth term in $\Delta \LL_{\cancel{\text{CP}},\,L}$, Eq.~\eqref{DeltaL-CP-odd-L}) and have completely been listed in  Ref.~\cite{Gavela:2014vra}. On the other hand, the symmetric counterpart extending the aforementioned set $\cS_{i,\,L}(h)$ and accounting for all the possible CP--violating pure gauge and gauge--$h$ interactions up to the $p^4$--operators is described by the complete linearly independent set of 16 operators $\cS_{i,\,R}(h)$ (second term in $\Delta \LL_{\cancel{\text{CP}},\,R}$ of Eq.~\eqref{DeltaL-CP-odd-R}), therefore in total one has 32 non-linear operators, among the which, 20 of them (10 $\cS_{i,\,L}(h)$ + 10 $\cS_{i,\,R}(h)$) had already been listed in  Refs.~\cite{Zhang:2007xy,Wang:2008nk}. In here 12 additional operators have been found (6 $\cS_{i,\,L}(h)$ + 6 $\cS_{i,\,R}(h)$) and naturally promoted by the symmetries of the model (together with the $P_{LR}$--symmetry), such that the whole tower of operators making up the basis $\{\cS_{i,\,L}(h),\,\cS_{i,\,R}(h)\}$ is given by:
\begin{widetext}
\beq
\begin{aligned}
&\cS_{1,\,\chi}(h)= \gchi\,g'\,\eps\,B^{\mu\nu}\,\Tr\Big(\TLchi \WWWchiu\Big)\,\cF_{1,\,\chi}(h)\,,
&&\cS_{9}(h)= i\,\gchi\,\eps\,\Tr\Big(\TLchi\,\WWchiu\Big)\Tr\Big(\TLchi\,\VLchirhou\Big)\partial^{\sigma}\cF_{9,\chi}(h)\,,\\[1.6mm] 
&\cS_{2,\chi}(h)= i\,g'\,\eps\,B^{\mu\nu}\,\Tr\Big(\TLchi\,\VLchirhou\Big)\,\partial^{\sigma}\cF_{2,\chi}(h)\,,
&&\brown{\cS_{10,\chi}(h)= i\,\Tr\Big(\VLchimuu\,\cD_\nu\VLchinuu\Big)\,\Tr\Big(\TLchi\,\VLchimud\Big)\,\cF_{10,\chi}(h)}\,, \\[1.6mm] 
&\cS_{3,\chi}(h)= i\,\gchi\,\eps\,\Tr\Big(\WWchiu\,\VLchirhou\Big)\,\partial^{\sigma}\cF_{3,\chi}(h)\,,
&&\brown{\cS_{11,\chi}(h)= i\,\Tr\Big(\TLchi\,\cD_\mu\VLchimuu\Big)\,\Tr\Big(\VLchinuu\,\VLchinud\Big)\,\cF_{11}(h)}\,,  \\[1.6mm] 
&\cS_{4,\chi}(h)= \gchi\,\Tr\Big(\WWchiu\,\VLchimud\Big)\Tr\Big(\TLchi\,\VLchinud\Big)\cF_{4,\chi}(h)\,,
&&\brown{\cS_{12,\chi}(h)= i\,\Tr\Big([\VLchimuu,\TLchi]\,\cD_\nu\VLchinuu\Big)\,\partial_{\mu}\cF_{12,\chi}(h)}\,, \\[1.6mm] 
&\cS_{5,\chi}(h)= i\,\Tr\Big(\VLchimuu\,\VLchinuu\Big)\Tr\Big(\TLchi\,\VLchimud\Big)\partial_{\nu}\cF_{5,\chi}(h)\,,
&&\brown{\cS_{13,\chi}(h)= i\,\Tr\Big(\TLchi\,\cD^\mu\VLchimud\Big)\,\partial^{\nu}\partial_{\nu}\cF_{13,\chi}(h)}\,, \\[1.6mm] 
&\cS_{6,\chi}(h)= i\,\Tr\Big(\VLchimuu\,\VLchimud\Big)\Tr\Big(\TLchi\,\VLchinuu\Big)\partial_{\nu}\cF_{6,\chi}(h)\,,
&&\brown{\cS_{14,\chi}(h)= i\,\Tr\Big(\TLchi\,\cD^\mu\VLchimud\Big)\,\partial^{\nu}\cF_{14,\chi}(h)\,\partial_{\nu}\cF'_{14,\chi}(h)} \,, \\[1.6mm] 
&\cS_{7,\chi}(h)= \gchi\, \Tr\Big(\TLchi\,\Big[\WWchiu,\VLchimud\Big]\Big)\,\partial_{\nu}\cF_{7,\chi}(h)\,,
&&\cS_{15,\chi}(h)= i\,\Tr\Big(\TLchi\,\VLchimuu\Big)\,\Big(\Tr\Big(\TLchi\,\VLchinuu\Big)\Big)^2\,\partial_{\mu}\cF_{15,\chi}(h)\,,\\[1.6mm] 
&\cS_{8,\chi}(h)= \gchi^2\,\eps\,\Tr\Big(\TLchi\,\WWchiu\Big)\Tr\Big(\TLchi\,\WWWchiu\Big)\cF_{8,\chi}(h)\,,
&&\brown{\cS_{16,\chi}(h)= i\,\Tr\Big(\TLchi\,\cD^\mu\VLchimud\Big)\,\Big(\Tr\Big(\TLchi\,\VLchinuu\Big)\Big)^2\,\cF_{16,\chi}(h)}\,.
\end{aligned}
\label{CP-odd-basis}
\eeq
\end{widetext}

\nt with $\WWchiu\equiv W^{\mu\nu,a}_\chi\tau^a/2$. In red color have been highlighted all those operators already listed in the context of CP--violating EW chiral effective theories coupled to a light Higgs in Ref.~\cite{Gavela:2014vra} and not provided in the left-right symmetric EW chiral treatment of Refs.~\cite{Zhang:2007xy,Wang:2008nk}. Notice that operators $\{\cS_{10-14,\,\chi}(h),\,\cS_{16,\,\chi}(h)\}$ containing the contraction $\cD_\mu \VLchimuu$, ($\cS_{13-14,\,\chi}(h)$  with double derivatives of $\cF(h)$), are not present in  Refs.~\cite{Zhang:2007xy,Wang:2008nk}, and are the resulting additional ones after extending the SM local symmetry through $SU(2)_R$, being naturally allowed by the local symmetries of the model. Notice as well that the entire basis $\cS_{i,\,R}(h)$ contained in Eq.~\eqref{CP-odd-basis} (for $\chi=R$) comes out just from the straightforward parity action under $P_{LR}$ of the operators tower $\cS_{i,\,L}$ in Ref.\cite{Gavela:2014vra}, or in other words, the whole basis $\cS_{i,\,R}(h)$ is mapped from $\cS_{i,\,L}(h)$ via $P_{LR}$--transformation (concerning the Goldstone boson part only, i.e before gauging the scenario).

It can be realized that the number of independent operators in the non--linear expansion turns out to be larger than for the analogous basis in the linear expansion, a generic feature when comparing both type of effective Lagrangians~\cite{Brivio:2013pma,Brivio:2014pfa}. The basis is also larger than that for the chiral expansions developed in the past for the case of a very heavy Higgs particle (i.e. absent at low energies)~\cite{Appelquist:1980vg,Longhitano:1980iz,Longhitano:1980tm,Appelquist:1993ka}, as:

\begin{itemize}

\item[i)] Terms which in the absence of the $\cF_i(h)$ functions were shown to be equivalent via total derivatives, are now independent.

\item[ii)] New terms including derivatives of $h$ appear.

\end{itemize}

The connection of the non-linear framework analysed in here to the effective linear scenarios explicitly implementing the SM Higgs doublet, has been done for the CP--conserving Lagrangian $\LL_\text{chiral} = \LL_0\,+\,\Delta \LL_{\text{CP},L}$ previously through~\cite{Alonso:2012px,Brivio:2013pma}, where all the corresponding non--linear CP--conserving operators were correspondingly weighted by powers of $\xi=v^2/\fL^2$, in order to keep track of their corresponding operator siblings in the linear side. Likewise, a similar connection to the effective linear scenarios has been done in~\cite{Gavela:2014vra} for the CP--violating non-linear Lagrangian $\LL_\text{chiral} = \LL_0
\,+\,\Delta \LL_{\cancel{\text{CP}},L}$, where each one of the operators in  the tower of Eq.~\eqref{CP-odd-basis} (for the case of $\chi=L$) were weighted as well by their corresponding powers of $\xi=v^2/\fL^2$. For the whole CP-violating Lagrangian in Eq.~\eqref{Lchiral} assumed in this work, such linking between both of the EFT sides would lead to account for the corresponding left--right symmetric extension of the effective linear approaches and it is beyond the scopes of this work.

Concerning the symmetry $P_{LR}$ mentioned in Section~\ref{Theoretical-framework}, in the context of a general effective $SO(5)/SO(4)$ composite Higgs model scenario~\cite{Contino:2011np}, the discrete parity $P_{LR}$ was shown to be an accidental symmetry up to $p^2$--order and broken by several $p^4$--operators. Exactly the same properties are shared by the non-linear EW bosonic $\cG$-invariant scenario studied recently in~\cite{Yepes:2015zoa}, as it is suspected from the fact that $SU(2)_L \otimes SU(2)_R \sim SO(4)$. Indeed, the corresponding LO $p^2$--Lagrangian analysed there in~\cite{Yepes:2015zoa}, explicitly exhibited $P_{LR}$ as an accidental symmetry of the approach (before the gauging). At higher momentum order in the Lagrangian expansion, the $p^4$--operators encoded in the corresponding $\Delta \LL_{\text{CP}}$ did not break $P_{LR}$ either. As soon as the $p^4$--operators made of mixed left and right--handed covariant structures were called in, non-zero contributions appeared to be triggering the breaking of $P_{LR}$ (see~\cite{Yepes:2015zoa} for more details).

Up to now all the possible CP non--invariant pure gauge and gauge--$h$ interactions allowed by the local $\cG$ symmetry have been encoded up to the $p^4$--non--linear operators contribution in the first three pieces of $\LL_\text{chiral}$ in Eq.~\eqref{Lchiral}, i.e. in 
$\LL_0\,+\,\LL_{0,LR}
\,+\,\Delta \LL_{\text{\cancel{CP}}}$ through Eqs.~\eqref{LLO}-\eqref{CP-odd-basis}. In the following section the $SU(2)_L-SU(2)_R$ interplay between both of the symmetries is faced by accounting for all the possible left--right symmetric CP--breaking interactions up to the $p^4$-order in the chiral Lagrangian $\LL_\text{chiral}$ and parametrized by the remaining piece in Eq.~\eqref{Lchiral}, i.e. $\Delta\LL_{\cancel{\text{CP}},LR}$.

%%%%%%%%%%%%%% 3.3 \Delta\LL_{\cancel{\text{CP}},LR} %%%%%%%%%%%%%%%

\boldmath
\subsection{$SU(2)_L-SU(2)_R$ interplay: $\Delta\LL_{\cancel{\text{CP}},LR}$}
\unboldmath

\nt The implementation of the covariant objects $\VLLRmuut$, $\TLLRt$ and $\WWLRut$ given in Eqs.~\eqref{Vtilde}, \eqref{Ttilde} and \eqref{Wtilde} respectively, allows to build up  the complete basis of independent CP--violating operators accounting for the mixing between $SU(2)_L$ and $SU(2)_R$ covariant structures, and can be encoded as
\be
\hspace*{-0.1cm}
\Delta\LL_{\text{\cancel{CP}},LR}=c_{W,LR}\,\cS_{W,LR}(h)\,\, \,+\,\,\,\sum_{i=3,\,i\neq 13,14}^{16} \hspace*{-0.4cm} c_{i(j),LR}\,\cS_{i(j),LR}(h)\,,
\label{DeltaL-CP-odd-LR}
\ee
\nt where the index $j$ spans over all the possible operators that can be built up from each $\cS_{i,\chi}(h)$ in Eq.~\eqref{CP-odd-basis}, and here labelled as $\cS_{i(j),LR}(h)$ (as well as their corresponding coefficients $c_{i(j),LR}$), whilst the first term in $\Delta\LL_{\cancel{\text{CP}},LR}$ encodes the operator
\be
\cS_{W,\,LR}(h)\,\, = -\dfrac{1}{2}\,\gL\,\gR\,\eps\Tr\left(\WWLut\,\WWWRut\right)\,\cF_{W,\,LR}(h)
\label{W-LR}
\ee
\nt The complete set of operators $\cS_{i(j),LR}(h)$ in the second term of $\Delta\LL_{\cancel{\text{CP}},LR}$ are listed as:

\begin{widetext}
\beq
\begin{aligned}
&\cS_{3(1)}(h)= i\,\gL\,\eps\,\Tr\left(\WWLut\,\VLRrhout\right)\,\partial^{\sigma}\cF_{3(1)}(h)\,,
&&\cS_{9(2)}(h)= i\,\gR\,\eps\,\Tr\left(\TLRt\,\WWRut\right)\Tr\left(\TLLt\,\VLLrhout\right)\,\partial^{\sigma}\cF_{9(2)}(h)\,,\\[1.mm]
&\cS_{3(2)}(h)= i\,\gR\,\eps\,\Tr\left(\WWRut\,\VLLrhout\right)\,\partial^{\sigma}\cF_{3(2)}(h)\,,
&&\brown{\cS_{10(1)}(h)= i\,\Tr\left(\VLLmuut\,\cD_\nu\VLLnuut\right)\,\Tr\left(\TLRt\,\VLRmudt\right)\,\cF_{10(1)}(h)}\,,\\
&\cS_{4(1)}(h)= \gL\,\Tr\left(\WWLut\,\VLLmudt\right)\Tr\left(\TLRt\,\VLRnudt\right)\cF_{4(1)}(h)\,,
&&\brown{\cS_{10(2)}(h)= i\,\Tr\left(\VLRmuut\,\cD_\nu\VLRnuut\right)\,\Tr\left(\TLLt\,\VLLmudt\right)\,\cF_{10(2)}(h)}\,,
\end{aligned}
\label{CP-odd-left-right-basis-I}
\eeq
\end{widetext}

\begin{widetext}
\beq
\hspace*{-1cm}
\begin{aligned}
&\cS_{4(2)}(h)= \gR\,\Tr\left(\WWRut\,\VLRmudt\right)\Tr\left(\TLLt\,\VLLnudt\right)\cF_{4(2)}(h)\,,
&&\brown{\cS_{10(3)}(h)= i\,\Tr\left(\VLLmuut\,\cD_\nu\VLRnuut\right)\,\Tr\left(\TLLt\,\VLLmudt\right)\,\cF_{10(3)}(h)}\,, \\[2mm]
&\cS_{4(3)}(h)= \gL\,\Tr\left(\WWLut\,\VLRmudt\right)\Tr\left(\TLRt\,\VLRnudt\right)\cF_{4(3)}(h)\,,
&&\brown{\cS_{10(4)}(h)= i\,\Tr\left(\VLRmuut\,\cD_\nu\VLLnuut\right)\,\Tr\left(\TLRt\,\VLRmudt\right)\,\cF_{10(4)}(h)}\,, \\[2mm]
&\cS_{4(4)}(h)= \gR\,\Tr\left(\WWRut\,\VLLmudt\right)\Tr\left(\TLLt\,\VLLnudt\right)\cF_{4(4)}(h)\,,
&&\brown{\cS_{10(5)}(h)= i\,\Tr\left(\VLLmuut\,\cD_\nu\VLRnuut\right)\,\Tr\left(\TLRt\,\VLRmudt\right)\,\cF_{10(5)}(h)}\,, \\[2mm]
&\cS_{4(5)}(h)= \gL\,\Tr\left(\WWLut\,\VLRmudt\right)\Tr\left(\TLLt\,\VLLnudt\right)\cF_{4(5)}(h)\,,
&&\brown{\cS_{10(6)}(h)= i\,\Tr\left(\VLRmuut\,\cD_\nu\VLLnuut\right)\,\Tr\left(\TLLt\,\VLLmudt\right)\,\cF_{10(6)}(h)}\,, \\[2mm]
&\cS_{4(6)}(h)= \gR\,\Tr\left(\WWRut\,\VLLmudt\right)\Tr\left(\TLRt\,\VLRnudt\right)\cF_{4(6)}(h)\,,
&&\brown{\cS_{11(1)}(h)= i\,\Tr\left(\TLLt\,\cD_\mu\VLLmuut\right)\,\Tr\left(\VLRnuut\,\VLRnudt\right)\,\cF_{11(1)}(h)}\,,  \\[2mm]
&\cS_{5(1)}(h)= i\,\Tr\left(\VLLmuut\,\VLLnuut\right)\Tr\left(\TLRt\,\VLRmudt\right)\partial_{\nu}\cF_{5(1)}(h)\,,
&&\brown{\cS_{11(2)}(h)= i\,\Tr\left(\TLRt\,\cD_\mu\VLRmuut\right)\,\Tr\left(\VLLnuut\,\VLLnudt\right)\,\cF_{11(2)}(h)}\,,  \\[2mm]
&\cS_{5(2)}(h)= i\,\Tr\left(\VLRmuut\,\VLRnuut\right)\Tr\left(\TLLt\,\VLLmudt\right)\partial_{\nu}\cF_{5(2)}(h)\,,
&&\brown{\cS_{11(3)}(h)= i\,\Tr\left(\TLLt\,\cD_\mu\VLLmuut\right)\,\Tr\left(\VLLnuut\,\VLRnudt\right)\,\cF_{11(3)}(h)}\,,  \\[2mm]
&\cS_{5(3)}(h)= i\,\Tr\left(\VLLmuut\,\VLRnuut\right)\Tr\left(\TLLt\,\VLLmudt\right)\partial_{\nu}\cF_{5(3)}(h)\,,
&&\brown{\cS_{11(4)}(h)= i\,\Tr\left(\TLRt\,\cD_\mu\VLRmuut\right)\,\Tr\left(\VLLnuut\,\VLRnudt\right)\,\cF_{11(4)}(h)}\,,  \\[2mm]
&\cS_{5(4)}(h)= i\,\Tr\left(\VLLmuut\,\VLRnuut\right)\Tr\left(\TLRt\,\VLRnudt\right)\partial_{\mu}\cF_{5(4)}(h)\,,
&&\brown{\cS_{12(1)}(h)= i\,\Tr\left([\VLLmuut,\TLLt]\,\cD_\nu\VLRnuut\right)\,\partial_{\mu}\cF_{12(1)}(h)}\,, \\[2mm] 
&\cS_{5(5)}(h)= i\,\Tr\left(\VLLmuut\,\VLRnuut\right)\Tr\left(\TLRt\,\VLRmudt\right)\partial_{\nu}\cF_{5(5)}(h)\,,
&&\brown{\cS_{12(2)}(h)= i\,\Tr\left([\VLRmuut,\TLRt]\,\cD_\nu\VLLnuut\right)\,\partial_{\mu}\cF_{12(2)}(h)}\,, \\[2mm] 
&\cS_{5(6)}(h)= i\,\Tr\left(\VLLmuut\,\VLRnuut\right)\Tr\left(\TLRt\,\VLRnudt\right)\partial_{\mu}\cF_{5(6)}(h)\,,
&&\cS_{15(1)}(h)= i\,\Tr\left(\TLLt\,\VLLmuut\right)\left(\Tr\left(\TLRt\,\VLRnuut\right)\right)^2\,\partial_{\mu}\cF_{15(1)}(h)\,,\\[2mm]
&\cS_{6(1)}(h)= i\Tr\left(\VLLmuut\,\VLLmudt\right)\Tr\left(\TLRt\,\VLRnuut\right)\partial_{\nu}\cF_{6(1)}(h)\,,
&&\cS_{15(2)}(h)= i\,\Tr\left(\TLRt\,\VLRmuut\right)\left(\Tr\left(\TLLt\,\VLLnuut\right)\right)^2\,\partial_{\mu}\cF_{15(2)}(h)\,,\\[2mm]
&\cS_{6(2)}(h)= i\Tr\left(\VLRmuut\,\VLRmudt\right)\Tr\left(\TLLt\,\VLLnuut\right)\partial_{\nu}\cF_{6(2)}(h)\,,
&&\cS_{15(3)}(h)= i\,\Tr\left(\TLLt\,\VLLmuut\right)\Tr\left(\TLLt\,\VLLnuut\right)\Tr\left(\TLRt\,\VLRnudt\right)\,\partial_{\mu}\cF_{15(3)}(h)\,,\\[2mm]
&\cS_{6(3)}(h)= i\Tr\left(\VLLmuut\,\VLRmudt\right)\Tr\left(\TLLt\,\VLLnuut\right)\partial_{\nu}\cF_{6(3)}(h)\,,
&&\cS_{15(4)}(h)= i\,\Tr\left(\TLRt\,\VLRmuut\right)\Tr\left(\TLLt\,\VLLnuut\right)\Tr\left(\TLRt\,\VLRnudt\right)\,\partial_{\mu}\cF_{15(4)}(h)\,,\\[2mm]
&\cS_{6(4)}(h)= i\Tr\left(\VLLmuut\,\VLRmudt\right)\Tr\left(\TLRt\,\VLRnuut\right)\partial_{\nu}\cF_{6(4)}(h)\,,
&&\brown{\cS_{16(1)}(h)= i\,\Tr\left(\TLLt\,\cD_\mu\VLLmuut\right)\,\left(\Tr\left(\TLRt\,\VLRnuut\right)\right)^2\,\cF_{16(1)}(h)}\,,\\[2mm]
&\cS_{7(1)}(h)= \gL\,\Tr\left(\TLLt\left[\WWLut,\VLRmudt\right]\right)\,\partial_{\nu}\cF_{7(1)}(h)\,,
&&\brown{\cS_{16(2)}(h)= i\,\Tr\left(\TLRt\,\cD_\mu\VLRmuut\right)\,\left(\Tr\left(\TLLt\,\VLLnuut\right)\right)^2\,\cF_{16(2)}(h)}\,,\\[2mm]
&\cS_{7(2)}(h)= \gR\,\Tr\left(\TLRt\left[\WWRut,\VLLmudt\right]\right)\,\partial_{\nu}\cF_{7(2)}(h)\,,
&&\brown{\cS_{16(3)}(h)= i\,\Tr\left(\TLLt\,\cD_\mu\VLLmuut\right)\,\Tr\left(\TLLt\,\VLLnuut\right)\Tr\left(\TLRt\,\VLRnudt\right)\,\cF_{16(3)}(h)}\,,\\[2mm]
&\cS_{8(1)}(h)= \gL\,\gR\,\eps\,\Tr\left(\TLLt\,\WWLut\right)\Tr\left(\TLRt\,\WWWRut\right)
\cF_{8(1)}(h)\,,
&&\brown{\cS_{16(4)}(h)= i\,\Tr\left(\TLRt\,\cD_\mu\VLRmuut\right)\,\Tr\left(\TLLt\,\VLLnuut\right)\Tr\left(\TLRt\,\VLRnudt\right)\,\cF_{16(4)}(h)}\,,\\[2mm]
&\cS_{9(1)}(h)= i\,\gL\,\eps\,\Tr\left(\TLLt\,\WWLut\right)\Tr\left(\TLRt\,\VLRrhout\right)\,\partial^{\sigma}\cF_{9(1)}(h)\,,
&&\\
\end{aligned}
\label{CP-odd-left-right-basis-II}
\eeq
\end{widetext}

\nt where the suffix $LR$ in all $\cS_{i(j),LR}(h)$ and their corresponding $\cF_{i(j),LR}(h)$ has been omitted as well. Among the total 43 operators $\cS_{i(j),LR}(h)$ listed in Eqs.~\eqref{CP-odd-left-right-basis-I}-\eqref{CP-odd-left-right-basis-II}, 16 operators (highlighted in red color again) are missing in the left-right symmetric EW chiral treatment of Refs.~\cite{Zhang:2007xy,Wang:2008nk}. Through the operators tower in Eqs.~\eqref{CP-odd-left-right-basis-I}-\eqref{CP-odd-left-right-basis-II} the definition for $\cD_\mu \VLLRmuut$ follows a similar one as in Eq.~\eqref{U-transforming-objects}
\be
\cD_\mu \VLLRmuut\equiv \UHLR^\dagger\,\cD_\mu \VLLRmuu\,\UHLR\,,
\label{DVtilde-covariant-derivative}
\ee

\begin{table}[h] 
\centering
\renewcommand{\arraystretch}{2.2}
\begin{tabular}{||c|c||} 
\hline \hline
\multicolumn{2}{||c||}{\bf{$P_{LR}$ symmetry}} \\[-3mm]
\multicolumn{2}{||c||}{$\Longleftrightarrow$} \\
\hline\hline
\quad$\cS_{5(1,3,5)}(h)$ \quad\quad & \quad$\cS_{5(2,4,6)}(h)$ \quad\quad\\
\quad$\cS_{6(1,3)}(h)$ \quad\quad & \quad$\cS_{6(2,4)}(h)$ \quad\quad\\
\quad$\cS_{10(1,3,5)}(h)$ \quad\quad & \quad$\cS_{10(2,4,6)}(h)$ 
\quad\quad\\
\quad$\cS_{11(1,3)}(h)$ \quad\quad & \quad$\cS_{11(2,4)}(h)$ 
\quad\quad\\
\quad$\cS_{12(1)}(h)$ \quad\quad & \quad$\cS_{12(2)}(h)$ \quad\quad\\
\quad$\cS_{15(1,3)}(h)$ \quad\quad & \quad$\cS_{15(2,4)}(h)$ \quad\quad\\
\quad$\cS_{16(1,3)}(h)$ \quad\quad & \quad$\cS_{16(2,4)}(h)$ \quad\quad\\[1mm]
\hline\hline
\end{tabular}
\caption{\sf $P_{LR}$--symmetry acting over a subset of operators from $\Delta\LL_{\cancel{\text{CP}},LR}$ in Eqs.~\eqref{CP-odd-left-right-basis-I}-\eqref{CP-odd-left-right-basis-II}. The rest of the operators from $\Delta\LL_{\cancel{\text{CP}},LR}$ not listed here are $P_{LR}$--even. For the CP--violating case, there are no operators explicitly breaking $P_{LR}$, although some non-linear CP--conserving $p^4$--operators will trigger its breaking~\cite{Yepes:2015zoa}.}
\label{PLR-tranformation-properties-odd}
\end{table}

\nt where $\DLL^\mu \VLLRnuu$ has been defined in Eq.~\eqref{DV-covariant-derivative}. Notice that operators $\cS_{10(1-6)}(h)$, $\cS_{11(1-4)}(h)$ and $\cS_{16(1-4)}(h)$ containing the contraction $\cD_\mu \VLLRmuu$, whereas $\cS_{12(1-2)}(h)$ involving one derivative of $\cF(h)$, are not present in~\cite{Zhang:2007xy,Wang:2008nk}, and are the resulting additional ones from the allowed $SU(2)_L$--$SU(2)_R$ interplay of $\Delta\LL_{\cancel{\text{CP}},LR}$, together with the $P_{LR}$--symmetry.

The interplaying CP--breaking non--linear  operators listed in Eqs.~\eqref{CP-odd-left-right-basis-I}-\eqref{CP-odd-left-right-basis-II}, can be catalogued as: 

\begin{itemize}

\item[a)] $\cS_{3(1-2)}(h)$, $\cS_{4(1-6)}(h)$, $\cS_{5(1-6)}(h)$, $\cS_{6(1-4)}(h)$, $\cS_{7(1-2)}(h)$, $\cS_{8(1)}(h)$ and $\cS_{9(1-2)}(h)$ coming from a direct extension of the original Appelquist-Longhitano chiral Higgsless basis already considered in~\cite{Appelquist:1980vg,Longhitano:1980iz,Longhitano:1980tm,
Feruglio:1992wf,Appelquist:1993ka} (for the CP--breaking sector), coupled to the light Higgs $\cF(h)$ insertions, and after applying the discrete parity $P_{LR}$.

\item[b)] $\cS_{10(1-6)}(h)$, $\cS_{11(1-4)}(h)$ and $\cS_{16(1-4)}(h)$ containing the contraction $\cD_\mu \VLchimuu$ and no derivative couplings from $\cF(h)$. 

\item[c)] $\cS_{3(1-2)}(h)$, $\cS_{5(1-6)}(h)$, $\cS_{6(1-4)}(h)$, $\cS_{7(1-2)}(h)$, $\cS_{9(1-2)}(h)$, $\cS_{12(1-2)}(h)$ and $\cS_{15(1-4)}(h)$ with one derivative coupling from $\cF(h)$.

\end{itemize}

\nt Finally, the transformation properties under the parity symmetry $P_{LR}$ of some of the operators from $\Delta\LL_{\cancel{\text{CP}},LR}$ in Eqs.~\eqref{CP-odd-left-right-basis-I}-\eqref{CP-odd-left-right-basis-II} are exhibited in the Table~\ref{PLR-tranformation-properties-odd}, with the operators not collected in there transforming as $P_{LR}$--even. Compact notation $\cS_{i(j,k,l),LR}(h)$ in the left column stands for operators $\cS_{i(j),LR}(h)$, $\cS_{i(k),LR}(h)$, $\cS_{i(l),LR}(h)$ reflected to $\cS_{i(m),LR}(h)$, $\cS_{i(n),LR}(h)$ $\cS_{i(p),LR}(h)$ (or vice versa) respectively, and collected by the operator notation $\cS_{i(m,n,p),LR}(h)$ in the right column. As it can be noticed, the $P_{LR}$ still holds as an accidental symmetry up tp $p^4$--contributions in the Lagrangian expansion. Conversely, some non-linear CP--conserving $p^4$--operators will trigger its breaking explicitly~\cite{Yepes:2015zoa}, as it was expected from the general composite Higgs models grounds~\cite{Contino:2011np,Alonso:2014wta}.

Some of the CP non--conserving bosonic operators provided above can be directly translated into pure bosonic operators  plus fermionic-bosonic ones. In fact, some of the operators in Eq.~\eqref{CP-odd-basis} (for the case of $\chi=L$) had not been explored, but traded instead by fermionic ones via
the equations of motion~\cite{Buchalla:2013rka}. Such connection can be established through the covariant derivative $\cD_\mu\,\VLchimuu$ and the corresponding equation of motion (EOM) for the light Higgs field, as it has been described in the non--linear left--right CP--conserving treatment of Ref.~\cite{Yepes:2015zoa}. Following the same reasoning line as in~\cite{Yepes:2015zoa}, it is inferred that 

\begin{itemize}

\item For the massless fermion case, operators $\{\cS_{10-12,\,R}(h),\,\cS_{14,\,R}(h),\,\cS_{16,\,R}(h)\}$ with the contraction $\cD_\mu\,\VLRmuu$ in Eq.~\eqref{CP-odd-basis}, can be traded by pure bosonic operators contained in $\Delta\LL_{\cancel{\text{CP}}}$ (Eq.~\eqref{CP-odd-basis}), some of them with the structure $\cD_\mu\,\VLLmuu$, and therefore they can be disregarded from the final operator basis in $\Delta\LL_{\cancel{\text{CP}}}$.  Similar feature applies for the operators $\cS_{10(1-6)}(h)$, $\cS_{11(1-4)}(h)$, $\cS_{12(1-2)}(h)$, and $\cS_{16(1-4)}(h)$ from $\Delta\LL_{\cancel{\text{CP}},LR}$ (Eqs.~\eqref{CP-odd-left-right-basis-I}-\eqref{CP-odd-left-right-basis-II}). In general, for the massive fermion case, all the previous operators have to be retained in the final basis.

\item For the vanishing fermion case, operator $\cS_{13,\,R}(h)$ with double derivatives of $\cF(h)$ in Eq.~\eqref{CP-odd-basis} is rewritable in terms of bosonic operators, with some of them contained in $\Delta\LL_{\cancel{\text{CP}}}$ and others in $\Delta\LL_{\cancel{\text{CP}},LR}$, and thus can be disregarded from the final operator basis. No operators from ~$\Delta\LL_{\cancel{\text{CP}},LR}$ (Eqs.~\eqref{CP-odd-left-right-basis-I}-\eqref{CP-odd-left-right-basis-II}) are trade-able to bosonic operators as there is no double derivative couplings of $\cF(h)$ in ~$\Delta\LL_{\cancel{\text{CP}},LR}$.  When fermion masses are switched on, $\cS_{13,\,R}(h)$ is physical and it has to be included in the final basis.

\end{itemize}

%%%%%%%%%%%%%%%%%  3.6 EOM and heavy RH fields decoupling %%%%%%%%%%%%%

\subsection{Integrating-out heavy right handed fields}
\label{Integrating-out}

\nt It is possible to integrate out the right handed gauge fields from the physical spectrum via the equations of motion for the strength gauge fields $W^{\mu,a}_R$ as it was done for the CP--conserving case in~\cite{Yepes:2015zoa,Shu:2015cxm}. A non-trivial EOM for the CP--violating case is obtained by including the analogous right handed counterparts for the strength kinetic and custodial conserving terms of $\LL_0$ in~\eqref{LLO}, together with the mixing effects from the term
$c_{C,LR}\,\cP_{C,LR}(h)$, with $c_{C,LR}$ the coefficient for the CP--conserving left--right operator $\cP_{C,\,LR}(h)\,\, = - \frac{1}{2}\,\fL\,\fR\,\Tr\Big(\VLLmuut\VLRmudt\Big)\,\cF_{C,\,LR}(h)$. By accounting for these contributions it is obtained then
\be
\VLRmuu\,\equiv\,-\epsilon\,\,\VLLmuu\,,\qquad
\text{with}\qquad\epsilon\equiv \frac{\fL}{\fR}\,c_{C,\,LR}\,.
\label{Gauge-field-EOM}
\ee

\nt The latter relation can be translated into the unitary gauge as
\be
\begin{aligned}
W_{\mu,\,R}^\pm &\quad\Rightarrow\quad -\frac{\gL}{\gR}\,\epsilon\,\,W_{\mu,\,L}^\pm\,,\\
W_{\mu ,R}^3&\quad\Rightarrow\quad \frac{g'}{\gR}\left(1+\epsilon\right)B_{\mu } -\frac{\gL}{\gR}\,\epsilon\,\,W_{\mu ,L}^3
\label{Gauge-field-EOM-unitary-gauge}
\end{aligned}
\ee

\nt By replacing the Eq.~\eqref{Gauge-field-EOM} through~\eqref{DeltaL-CP-odd-R}-\eqref{CP-odd-basis} (for $\chi=R$) and~\eqref{DeltaL-CP-odd-LR}-\eqref{CP-odd-left-right-basis-II}, all  the right handed and left-right operators will collapse onto the left  ones, affecting the corresponding global coefficients $c_{i,L}$ in a generic manner as
\be
c_{i,L}\,\,\Longrightarrow \,\,\tilde{c}_{i,L} \,=\, c_{i,L} \,+\, \sum^{4}_{k=1}\epsilon^k\,\cF^{(k)}\left(c_{i,R},\,c_{i(j)},\,c_{l(m)}\right) 
\label{Coefficients-redefined}
\ee

\nt where the functions $\cF^{(k)}\left(c_{i,R},c_{i(j)},\,c_{l(m)}\right)$ will encode linear combinations on the coefficients $c_{i,R}$, $c_{i(j)}$ and additional mixing left-right operators via $c_{l(m)}$. The number of fields $\VLRmuu$ through each one of the right and left--right operators determines the $\fL/\fR$--suppression for the contributions induced onto the left handed operators. Consequently, in the limiting case $\fL\ll \fR$ at low energies, it is realized that the set of non-linear operators 
\be
\{\cS_B(h),\,\cS_{2D,L}(h),\,\cS_{2,L}(h)\}\,
\label{Sensitive-operators}
\ee

\nt is sensitive to the contributions, up to the order $\cO(\epsilon)$, from the right handed operators
\be
\{\cS_{W,R}(h),\,\cS_{1,R}(h),\,\cS_{8,R}(h)\}\,
\label{Right-operators}
\ee

\nt and the mixing left--right set
\be
\{\cS_{2D,LR}(h),\,\cS_{\text{9(2)}}(h)\}\,.
\label{Left-Right-operators}
\ee  

\nt It can also be realized that the CP--violating self-couplings of the electroweak gauge bosons will be sensitive only to the left handed operators at low energies. In fact, following Ref.~\cite{Hagiwara:1986vm}, the CP-odd sector of the Lagrangian that
describes triple gauge boson vertices (TGVs) can be parametrised as 
\begin{widetext}
\begin{eqnarray} 
{\mathcal L}^{WWV}_{\text{eff, \cancel{CP}}} = g_{WWV} \Bigg( &
g_4^V W^\dagger_\mu W_\nu (\partial^\mu V^\nu + \partial^\nu V^\mu)
 - i \tilde{\kappa}_V W^\dagger_\mu W_\nu \tilde{V}^{\mu\nu}
- i \frac{\tilde{\lambda}_V}{M_W^2} W^\dagger_{\sigma\mu} W^\mu_\nu 
\tilde{V}^{\nu\sigma}\nonumber\\
&+  \tilde{g}_{6}^V ( W^\dagger_\nu\partial_\mu W^\mu + W_\nu \partial_\mu W^{\dagger\mu} )V^\nu
+  \tilde{g}_{7}^V W^\dagger_\mu W^\mu \partial^\nu V_\nu
\Bigg)\,, 
\label{eq:wwv}
\end{eqnarray}
\end{widetext}

\nt where $V \equiv \{\gamma, Z\}$ and $g_{WW\gamma} \equiv e$, $g_{WWZ} \equiv e\,c_W/s_W$, with $W^\pm_{\mu\nu}$ and $V_{\mu\nu}$ standing for the kinetic part of the implied gauge field strengths. The dual field-tensor of any field strength $V_{\mu\nu}$ is defined as $\tilde{V}^{\mu\nu} \equiv \frac{1}{2} \epsilon^{\mu\nu\rho\sigma}V_{\rho\sigma}$. The compact notation $c_W\equiv \cos\theta_W$ and $s_W\equiv \sin\theta_W$ is implicit, with $\theta_W$ the Weinberg angle. In writing Eq.~(\ref{eq:wwv}) we have introduced the
coefficients $\tilde g_{6}^V$ and $\tilde g_7^V$ associated to operators that contain the contraction $\DLL_\mu\VL^\mu$; its $\partial_\mu\VL^\mu$ part vanishes only for on-shell gauge bosons; in all generality $\DLL_\mu\VL^\mu$ insertions could only be disregarded in the present context when fermion masses are neglected.
In the SM all couplings in Eq.~(\ref{eq:wwv}) vanish.

Electromagnetic gauge invariance requires $g_{4}^{\gamma} =0$, while $\tilde{g}_{6}^\gamma  =\tilde{g}_{7}^\gamma=\tilde \lambda_\gamma=\tilde\lambda_Z=0$. All the other effective couplings in~\eqref{eq:wwv} are given by
\be
\begin{aligned}
 \tilde{\kappa}_\gamma &=-\frac{4e^2}{s_W^2}\left(c_{1,L}+ 2\,c_{8,L}\right), \hspace*{3mm} 
\tilde{\kappa}_Z =\frac{4e^2}{c_W^2}
\left(c_{1,L}- 2\frac{c_W^2}{s_W^2}\,c_{8,L}\right) \\
g_4^Z &=\frac{e^2}{2c_W^2s_W^2}\,c_{4,L}, 
\hspace*{16mm} \tilde{g}_{6}^Z =\frac{e^2}{2c_W^2s_W^2}\left(c_{4,L} + c_{10,L} \right),\\
\tilde{g}_{7}^Z &=-\frac{e^2}{2c_W^2s_W^2}\left(c_{4,L} - 2\,c_{11,L} \right)\,.
\label{eq:coeftgv}
\end{aligned}
\ee

\nt An additional contribution to the $ZZZ$ vertex arises out from the operators $\cS_{10-11,L}(h)$ and $\cS_{6,L}(h)$ as
\begin{equation} 
{\mathcal L}^{3Z}_{\text{eff, \cancel{CP}}}\, = 
\tilde g_{3Z} \, Z_\mu Z^\mu \partial_\nu Z^\nu \, , 
\label{eq:zzz}
\end{equation}

\nt with 
\begin{equation}
\tilde g_{3Z}=\frac{e^3}{2c^3_W\,s^3_W} 
\left( c_{10,L} + c_{11,L} + 2c_{16,L} \right) \, ,
\label{eq:g3z}
\end{equation}

\nt which, alike to the phenomenological couplings 
$\tilde g_{6}^V$ and $\tilde g_7^V$ in Eq.~(\ref{eq:wwv}),
vanishes for on-shell $Z$ bosons and in general can be disregarded in
the present context when the masses of fermions coupling to the $Z$
are neglected. 
%%%%%%%%%%%%%%%%%%%%%%%%%%%%%%%%%%%%%%%%%%%%%%%%%%%%%%%%%%%%%%%%%%%%%%%

The coupling $ \tilde{\kappa}_\gamma$ induces one-loop contributions to the fermion electric dipole moments (EDMs), the which are generically the best windows on BSM sources of CP-violation, due to the
combination of the very stringent experimental bounds with the fact
that they tend to be almost free from SM background contributions. The amplitude corresponding to the one-loop fermionic EDM can be parametrised as
\begin{align}
\mathcal{A}_{f}&\equiv -i\,d_f\,\overline{u}\left(p_2\right)\,\sigma_{\mu \nu}q^{\nu }\gamma^5 \,u\left(p_1\right)\,,
\end{align}

\nt where $d_f$ denotes the fermionic EDM strength. The 1--loop
integral diverges logarithmically\footnote{For a specific UV model which does not lead to logarithmic diverging EDM see \cite{Appelquist:2004mn}.}; assuming a physical cut-off
$\Lambda_s$ for the high energy BSM theory, following the generic
computation in Ref.~\cite{Marciano:1986eh} and implementing the coupling $ \tilde{\kappa}_\gamma$ in~\eqref{eq:coeftgv}, we obtain the EDM coefficient
\be
d_f=\left(c_{1,L} + 2\,c_{8,L}\right)\frac{e^3\,G_F\,T_{3 L}}{\sqrt{2}\,\pi ^2\,s^2_W}\,\,m_f\,\left[\log \left(\frac{\Lambda_s
    ^2}{M_W^2}\right)+ \mathcal{O}(1)\right]\,,
\label{EDM-coefficient}
\ee

\nt where $T_{3L}$ stands for the fermion weak isospin and $G_F$ the Fermi coupling constant.
The present experimental bound on the electron  EDM~\cite{Baron:2013eja} $\left|d_e/e\right|<8.7\times 10^{-29}\,\text{cm}$ at 90\% CL, and the corresponding one for the neutron~\cite{Baker:2006ts} $\left|d_n/e\right|<2.9\times 10^{-26}\,\text{cm}$ at 90\% CL\footnote{Constituent quark masses $m_u=m_d=m_N/3$ have been used.}, entails then a limit\footnote{Weaker but more direct bounds on these operators can be imposed
from the study of $W\gamma$ production at colliders. For example, in Ref.~\cite{Dawson:2013owa} it was concluded that the future 14 TeV LHC data with 10 fb$^{-1}$ can place  a 95\% CL bound
\begin{equation}
|\tilde\kappa_\gamma|\leq 0.05\quad \Longrightarrow \quad |c_{1,L}+2 c_{8,L}|\leq 0.03 \, .
\label{eq:wwA}
\end{equation}} 
\be
\left|\left(c_{1,L} + 2\,c_{8,L}\right)\left[\log \left(\frac{\Lambda_s
    ^2}{M_W^2}\right) + \mathcal{O}(1)\right]\right|<5.2 (2.8)\times 10^{-5}\,.
\label{c1-c8-e-nEDM}
\ee

\nt Direct constraints on CP-violating effects
in the $WWZ$ vertex can be imposed by combining the results using
the LEP collaboration studies on the observation of the angular
distribution of $W's$ and their decay products in $WW$ production at
LEPII~\cite{Abbiendi:2000ei,Abdallah:2008sf,Schael:2004tq}. Such combination yields the 1$\sigma$ (68\% CL) constraints~\cite{Gavela:2014vra}
\beq
\hspace*{-1mm}
-1.8 \leq c_{4,L} \leq -0.50 \,,\quad
-0.29\leq c_{1,L} - 2\,\frac{c^2_W}{s^2_W}\,c_{8,L}\leq -0.13\,.
\label{eq:wwzlep2}
\eeq

\nt Effective CP--violating $hVV$--couplings turns out to be also affected in the framework. In particular the vertex $hZZ$ can be parametrized as
\beq
{\cal L}_{{\rm eff, \cancel{CP}}}^{hVV}\,\,\supset\,\,\tilde{g}_{H Z Z}  \, h\,Z_{\mu \nu}\,\tilde{Z}^{\mu \nu} \,,
\label{eq:hZZ}
\eeq

\nt with a tree level contribution
\beq
\hspace*{-2mm}
\begin{aligned}
\frac{\fL\,c_W^2}{4e^2s_W^2}\,\tilde{g}_{H ZZ}=&-\frac{1}{4}\hat{a}_{\tilde{B}} +
\frac{c_W^4}{s_W^4}\,\hat{a}_{8,L}-\frac{c_W^2}{s_W^2}
\,\hat{a}_{1,L}+ \frac{1}{2s_W^2}\,\hat{a}_{2,L}+\\
& -\frac{c_W^4}{8s_W^4}\,\hat{a}_{W,L}-\frac{c_W^2}{2s_W^4}
\,\hat{a}_{9,L}-\frac{c_W^2}{4s_W^4}\,\hat{a}_{3,L}
\label{eq:coefhZZ}
\end{aligned}
\eeq

\nt where the coefficients $\hat{a_i}$ are defined for simplicity as $
\hat{a}_i \equiv c_i a_i$, with $a_i$ the coefficients from the $\cF(h)$--definition in~\eqref{F}. The coupling $\tilde{g}_{H ZZ}$ is useful in parametrising the decay $h\rightarrow ZZ$ as~\cite{Chatrchyan:2013mxa,Gao:2010qx}. In Ref.~\cite{Chatrchyan:2013mxa} a measure of  CP-violation in the
decay $h\rightarrow ZZ^*\rightarrow 4\,l$ was defined as
\begin{eqnarray}
 f_{d_2}=\frac{|d_2|^2\sigma_2}{|d_1|^2\sigma_1+|d_2|^2\sigma_2}
\end{eqnarray}

\nt where
\be
d_1=2\,i\,,\qquad d_2=-2\,i\,\fL\,\tilde g_{HZZ}\, ,
\ee

\nt with $\fL$ the EW scale and $\sigma_1$ ($\sigma_2$) the corresponding cross section for the process $h\rightarrow ZZ$ when $d_1=1$ ($d_2=1$) and $d_2=0$ ($d_1=1$).
For $M_h=125.6$ GeV, $\frac{\sigma_1}{\sigma_2}=6.36$. In
Ref.~\cite{Chatrchyan:2013mxa}
$f_{d_2}$ was fitted as one of the parameters of the multivariable analysis, obtaining the measured value $
 f_{d_2}=0.00^{+0.17}_{-0.00}$, implying then $\frac{|d_2|}{|d_1|}=0.00^{+1.14}_{-0.00}$ and pointing to the CP-even nature of the state. Furthermore, $95\%$ CL exclusion bounds on $f_{d_2}$ were derived as $f_{d_2}<0.51$, entailing thus $\frac{|d_2|}{|d_1|}<2.57$. We can directly translate these bounds to 68\,(95)\% CL constraints on the coefficients of the relevant CP-violating operators through the coupling in~\eqref{eq:coefhZZ} as
\begin{equation}
\frac{\fL\,c_W^2}{4\,e^2\,s_W^2}\,\tilde{g}_{H ZZ}\leq 10.3\ (23.3)\,. 
\label{tired3}
\end{equation}

\nt None of the involve coefficients through the previous couplings receive contributions at low energies from the right handed operators nor the left--right ones. For a high energy scale $\fR$ not far above the EW scale $\fL$, additional operators would contribute onto the left handed ones as the ratio $\fL/\fR$ would be non-negligible. Nonetheless  , these additional contributions turn out to be small as the small allowed range $-0.02<c_{C,\,LR}<0.02$~\cite{Shu:2015cxm} suppresses the scale ratio and therefore the parameter $\epsilon$ in~\eqref{Gauge-field-EOM}. For the hypothetical case of $c_{C,\,LR}\sim 1$ and $\fR \approx \fL$, the right and left--right operators contribution are enhanced, and all the coefficients through the couplings in~\eqref{eq:coeftgv}, \eqref{eq:g3z}, \eqref{c1-c8-e-nEDM}, \eqref{eq:wwzlep2} and \eqref{eq:coefhZZ} become
\be
\begin{aligned}
\tilde{c}_{B}&=c_{B}\,+\,\epsilon \Big[2\,c_{W,R}-4\,c_{8,R}-4\left(c_{1,R}+c_{8,R}\right)\Big]\,,
\\
\tilde{c}_{W,L}&=c_{W,L}\,-\,2\,\epsilon\,c_{W,LR}\,,
\\
\tilde{c}_{1,L}&=c_{1,L}\,+\,\frac{\epsilon}{4}\Big[c_{W,R}-4 \left(c_{1,R}+c_{8,R}\right)\Big]\,,
\\
\tilde{c}_{2,L}&=c_{2,L}\,-\,\epsilon \left[\frac{\left(a_{2,R} \,c_{2,R}+a_{9,R}\,c_{9,R}\right)}{a_{2,L}}\,-\,\frac{a_{\text{9(2)}} \,c_{\text{9(2)}}}{a_{2,L}}\right]\,,
\\
\tilde{c}_{3,L}&=c_{3,L}\,-\,\epsilon\,\frac{a_{\text{3(1)}} \,c_{\text{3(1)}}+a_{\text{3(2)}}\,c_{\text{3(2)}}}{2 a_{3,L}}\,+\, 
\\
&\phantom{=c_{3,L}}\,-\,\epsilon\,\frac{a_{\text{9(1)}} \,c_{\text{9(1)}}+a_{\text{9(2)}}\,c_{\text{9(2)}}}{ a_{3,L}}\,,
\\
\tilde{c}_{4,L}&=c_{4,L}\,-\,\frac{\epsilon}{2}\left(c_{\text{4(1)}}+c_{\text{4(4)}}+c_{\text{4(5)}}\right)\,,
\\
\tilde{c}_{8,L}&=c_{8,L}\,-\,\epsilon\,c_{\text{8(1)}}\,,
\\
\tilde{c}_{9,L}&=c_{9,L}\,-\,\epsilon\,\frac{a_{\text{9(1)}} \,c_{\text{9(1)}}+a_{\text{9(2)}}\,c_{\text{9(2)}}}{a_{9,L}}\,,
\\
\tilde{c}_{10,L}&=c_{10,L}\,-\,\epsilon\left(c_{\text{10(1)}}+c_{\text{10(3)}}+c_{\text{10(6)}}\right)\,,
\\
\tilde{c}_{11,L}&=c_{11,L}\,-\,\epsilon \left(c_{\text{11(2)}}+c_{\text{11(3)}}\right)\,,
\\
\tilde{c}_{16,L}&=c_{16,L}\,-\,\epsilon \left(c_{\text{16(2)}}+c_{\text{16(3)}}\right)\,.
\label{Modified-coefficients}
\end{aligned}
\ee

\nt Counting the number of right handed and left--right operators appearing through the coefficients in~\eqref{Modified-coefficients},  lead us to have finally an effective set of 40 operators in total\,\,=\,\,20 left ops. (set in Eq.~\eqref{G-2D}+Eq.~\eqref{CP-odd-basis})\,\,+\,\,5 right ops. (in Eq.~\eqref{Modified-coefficients})\,\,+\,\,15 left--right ops (in Eq.~\eqref{Modified-coefficients}). A right handed gauge sector far above the EW scale will imply a hierarchical case with NP effects parametrized via a much smaller operator basis as the $\fL/\fR$--suppression would entail, and leaving us therefore with 25 operators in total\,\,=\,\,20 left ops.\,\,+\,\,3 right ops. (in Eq.~\eqref{Right-operators})\,\,+\,\,2 left--right ops (in Eq.~\eqref{Left-Right-operators}).

%%%%%%%%%%%%%%%%%%%%%%   4.  Conclusions     %%%%%%%%%%%%%%%%%%%%%%

\section{Conclusions}
\label{Conclusions}

\nt The electroweak interactions taking place in our nature have exhibited a non--exact product of the charge conjugation and parity symmetries of particle physics. Moreover, the observed matter--antimatter asymmetry of our universe compel us
for considering new sources of CP--violation, that are signalled as well by the extreme fine-tuning entailed by the strong CP problem. 

Low energy effects from a CP--violating sector may be sourced from a 
NP field content playing a role at high energy regimes, reachable at the LHC, future facilities and colliders. An effective approach is in order thus to parametrize all those effects. In this paper, and concerning only the bosonic gauge sector, the NP field content is dictated by the existence of a spin--1 resonance sourced by the extension of the SM local gauge symmetry $\cG_{SM}=SU(2)_L\otimes U(1)_Y$ up to the larger local group 
$\cG=SU(2)_L\otimes SU(2)_R\otimes U(1)_{B-L}$, here described via
a non--linear EW scenario with a light dynamical Higgs, up to the $p^4$-contributions in the Lagrangian expansion, and focused only on the CP--violating sector.

This paper completes the CP--violating pure gauge and gauge-$h$ operator basis given in Refs.~\cite{Zhang:2007xy,Wang:2008nk} (for the CP--breaking sector) in the context of left--right symmetric EW chiral models, completing and generalizing as well the work done in Refs.~\cite{Appelquist:1980vg,Longhitano:1980iz,Longhitano:1980tm,
Feruglio:1992wf,Appelquist:1993ka} (for the CP--violating sector) with a heavy Higgs chiral scenario, and extending as well Ref.~\cite{Gavela:2014vra} for the CP--violating light Higgs dynamical framework, to the case of a larger local gauge symmetry $\cG$ in the context of non-linear EW interactions coupled to a light Higgs particle.

The work done in here may be considered as well as a generic UV completion of the low energy non-linear approaches of Refs.~\cite{Appelquist:1980vg,Longhitano:1980iz,Longhitano:1980tm,
Feruglio:1992wf,Appelquist:1993ka} (for the CP--breaking sector)  and  Ref.~\cite{Gavela:2014vra}, assuming the extended gauge field sector arising out from an energy regime higher than the EW scale. The physical effects induced by integrating out the right handed fields from the physical spectrum are analysed, in particular, the effects on the CP--violating gauge couplings (Eqs.~\eqref{eq:wwv}-\eqref{eq:g3z}, \eqref{eq:wwA} and~\eqref{eq:wwzlep2}), EDM observables (Eqs.~\eqref{EDM-coefficient} and~\eqref{c1-c8-e-nEDM}), effective coupling $hZZ$ and CP-violation in the
decay $h\rightarrow ZZ^*\rightarrow 4\,l$ (Eqs.~\eqref{eq:hZZ}-\eqref{eq:coefhZZ}). The relevant set of operators have been identified at low energies, 25 operators in total\,\,=\,\,20 left ops.\,\,+\,\,3 right ops. (in Eq.~\eqref{Right-operators})\,\,+\,\,2 left--right ops (in Eq.~\eqref{Left-Right-operators}). More low energy effects from a higher energy gauge sector~\cite{Shu:2015cxm,Shu:2016exh} could shed some light on the underlying NP playing role in our nature, and likely will aid us in understanding better the origin of the electroweak symmetry breaking mechanism.

%%%%%%%%%%%%%%%%%%%%%%%%%%%%%%%%%%%%%%%%%%%%%%%%%%%%%%%%%%%%%%%%
% Acknowledgements
%%%%%%%%%%%%%%%%%%%%%%%%%%%%%%%%%%%%%%%%%%%%%%%%%%%%%%%%%%%%%%%%
\section*{Acknowledgements}

\nt The authors of this paper acknowledges valuable comments from R.~Alonso, I.~Brivio, M.~B.~Gavela, J.~Gonzalez-Fraile, D.~Marzocca, L. Merlo and J. Sanz-Cillero. J.~Y and R.~K acknowledge KITPC financial support during the completion of this work.

%%%%%%%%%%%%%%%%%%%%%%%%%%%%%%%%%%%%%%%%%%%%%%%%%%%%%%%%%%%%%%%%%
%%%%%%%%%%%%%%%%%%%%%%%%%  Bibliography     
%%%%%%%%%%%%%%%%%%%%%%%%%%%%%%%%%%%%%%%%%%%%%%%%%%%%%%%%%%%%%%%%%
%
%\bibliography{biblio}{}

\begin{thebibliography}{10}

\bibitem{:2012gk}
{\bf ATLAS Collaboration} Collaboration, G.~Aad {\em et.~al.},  Phys.Lett. {\bf
  B716} (2012) 1--29, [\href{http://xxx.lanl.gov/abs/1207.7214}{{\tt
  arXiv:1207.7214}}].

\bibitem{:2012gu}
{\bf CMS Collaboration} Collaboration, S.~Chatrchyan {\em et.~al.},  Phys.Lett.
  {\bf B716} (2012) 30--61, [\href{http://xxx.lanl.gov/abs/1207.7235}{{\tt
  arXiv:1207.7235}}].


\bibitem{Englert:1964et}
F.~Englert and R.~Brout,  Phys.Rev.Lett. {\bf 13} (1964) 321--323.

\bibitem{Higgs:1964ia}
P.~W. Higgs,  Phys.Lett. {\bf 12} (1964) 132--133.

\bibitem{Higgs:1964pj}
P.~W. Higgs,  Phys.Rev.Lett. {\bf 13} (1964) 508--509.


\bibitem{Dell'Aquila:1985ve}
J.~R. Dell'Aquila and C.~A. Nelson, Phys.Rev. {\bf D33} (1986) 80.

\bibitem{Dell'Aquila:1985vc}
J.~R. Dell'Aquila and C.~A. Nelson, Phys.Rev. {\bf D33} (1986) 93.

\bibitem{Dell'Aquila:1985vb}
J.~R. Dell'Aquila and C.~A. Nelson, Phys.Rev. {\bf D33} (1986) 101.

\bibitem{Soni:1993jc}
A.~Soni and R.~Xu, Phys.Rev. {\bf D48} (1993) 5259--5263,
  [\href{http://xxx.lanl.gov/abs/hep-ph/9301225}{{\tt hep-ph/9301225}}].

\bibitem{Chang:1993jy}
D.~Chang, W.-Y. Keung, and I.~Phillips, Phys.Rev. {\bf D48}
  (1993) 3225--3234, [\href{http://xxx.lanl.gov/abs/hep-ph/9303226}{{\tt
  hep-ph/9303226}}].

\bibitem{Arens:1994wd}
T.~Arens and L.~Sehgal, Z.Phys. {\bf C66} (1995) 89--94,
  [\href{http://xxx.lanl.gov/abs/hep-ph/9409396}{{\tt hep-ph/9409396}}].

\bibitem{Choi:2002jk}
S.~Choi, D.~Miller, M.~Muhlleitner, and P.~Zerwas, Phys.Lett. {\bf B553} (2003) 61--71,
  [\href{http://xxx.lanl.gov/abs/hep-ph/0210077}{{\tt hep-ph/0210077}}].

\bibitem{Buszello:2002uu}
C.~Buszello, I.~Fleck, P.~Marquard, and J.~van~der Bij, Eur.Phys.J. {\bf C32} (2004) 209--219,
  [\href{http://xxx.lanl.gov/abs/hep-ph/0212396}{{\tt hep-ph/0212396}}].

\bibitem{Godbole:2007cn}
R.~M. Godbole, D.~Miller, and M.~M. Muhlleitner, JHEP {\bf 0712} (2007) 031,
  [\href{http://xxx.lanl.gov/abs/0708.0458}{{\tt arXiv:0708.0458}}].

\bibitem{Cao:2009ah}
Q.-H. Cao, C.~Jackson, W.-Y. Keung, I.~Low, and J.~Shu, Phys.Rev.
  {\bf D81} (2010) 015010, [\href{http://xxx.lanl.gov/abs/0911.3398}{{\tt
  arXiv:0911.3398}}].

\bibitem{Gao:2010qx}
Y.~Gao, A.~V. Gritsan, Z.~Guo, K.~Melnikov, M.~Schulze, {\em et.~al.}, Phys.Rev. {\bf D81} (2010) 075022,
  [\href{http://xxx.lanl.gov/abs/1001.3396}{{\tt arXiv:1001.3396}}].

\bibitem{DeRujula:2010ys}
A.~De~Rujula, J.~Lykken, M.~Pierini, C.~Rogan, and M.~Spiropulu, Phys.Rev. {\bf D82} (2010) 013003,
  [\href{http://xxx.lanl.gov/abs/1001.5300}{{\tt arXiv:1001.5300}}].

\bibitem{Plehn:2001nj}
T.~Plehn, D.~L. Rainwater, and D.~Zeppenfeld, Phys.Rev.Lett. {\bf 88} (2002) 051801,
  [\href{http://xxx.lanl.gov/abs/hep-ph/0105325}{{\tt hep-ph/0105325}}].

\bibitem{Buszello:2006hf}
C.~Buszello and P.~Marquard, \href{http://xxx.lanl.gov/abs/hep-ph/0603209}{{\tt
  hep-ph/0603209}}.

\bibitem{Hankele:2006ma}
V.~Hankele, G.~Klamke, D.~Zeppenfeld, and T.~Figy, Phys.Rev. {\bf D74}
  (2006) 095001, [\href{http://xxx.lanl.gov/abs/hep-ph/0609075}{{\tt
  hep-ph/0609075}}].

\bibitem{Klamke:2007cu}
G.~Klamke and D.~Zeppenfeld, JHEP {\bf 0704} (2007) 052,
  [\href{http://xxx.lanl.gov/abs/hep-ph/0703202}{{\tt hep-ph/0703202}}].

\bibitem{Englert:2012ct}
C.~Englert, M.~Spannowsky, and M.~Takeuchi, JHEP {\bf 1206} (2012) 108,
  [\href{http://xxx.lanl.gov/abs/1203.5788}{{\tt arXiv:1203.5788}}].

\bibitem{Odagiri:2002nd}
K.~Odagiri, JHEP {\bf 0303} (2003) 009,
  [\href{http://xxx.lanl.gov/abs/hep-ph/0212215}{{\tt hep-ph/0212215}}].

\bibitem{DelDuca:2006hk}
V.~Del~Duca, G.~Klamke, D.~Zeppenfeld, M.~L. Mangano, M.~Moretti, {\em
  et.~al.}, JHEP {\bf 0610} (2006) 016,
  [\href{http://xxx.lanl.gov/abs/hep-ph/0608158}{{\tt hep-ph/0608158}}].

\bibitem{Andersen:2010zx}
J.~R. Andersen, K.~Arnold, and D.~Zeppenfeld, JHEP {\bf 1006} (2010)
  091, [\href{http://xxx.lanl.gov/abs/1001.3822}{{\tt arXiv:1001.3822}}].

\bibitem{Englert:2012xt}
C.~Englert, D.~Goncalves-Netto, K.~Mawatari, and T.~Plehn, JHEP {\bf 1301} (2013) 148,
  [\href{http://xxx.lanl.gov/abs/1212.0843}{{\tt arXiv:1212.0843}}].

\bibitem{Djouadi:2013yb}
A.~Djouadi, R.~Godbole, B.~Mellado, and K.~Mohan, Phys.Lett.
  {\bf B723} (2013) 307--313, [\href{http://xxx.lanl.gov/abs/1301.4965}{{\tt
  arXiv:1301.4965}}].


\bibitem{Dolan:2014upa}
  M.~J.~Dolan, P.~Harris, M.~Jankowiak and M.~Spannowsky,
  Phys.\ Rev.\ D {\bf 90} (2014) 7,  073008, \href{http://xxx.lanl.gov/abs/1406.3322}{{\tt arXiv:1406.3322}}.


\bibitem{Christensen:2010pf}
N.~D. Christensen, T.~Han, and Y.~Li, Phys.Lett. {\bf B693} (2010) 28--35,
  [\href{http://xxx.lanl.gov/abs/1005.5393}{{\tt arXiv:1005.5393}}].

\bibitem{Desai:2011yj}
N.~Desai, D.~K. Ghosh, and B.~Mukhopadhyaya, Phys.Rev. {\bf D83} (2011) 113004,
  [\href{http://xxx.lanl.gov/abs/1104.3327}{{\tt arXiv:1104.3327}}].

\bibitem{Ellis:2012xd}
J.~Ellis, D.~S. Hwang, V.~Sanz, and T.~You, JHEP {\bf 1211} (2012) 134,
  [\href{http://xxx.lanl.gov/abs/1208.6002}{{\tt arXiv:1208.6002}}].

\bibitem{Godbole:2013saa}
R.~Godbole, D.~J. Miller, K.~Mohan, and C.~D. White, Phys.Lett.
  {\bf B730} (2014) 275--279, [\href{http://xxx.lanl.gov/abs/1306.2573}{{\tt
  arXiv:1306.2573}}].

\bibitem{Delaunay:2013npa}
C.~Delaunay, G.~Perez, H.~de~Sandes, and W.~Skiba, Phys.Rev. {\bf D89} (2014) 035004,
  [\href{http://xxx.lanl.gov/abs/1308.4930}{{\tt arXiv:1308.4930}}].

\bibitem{Voloshin:2012tv}
M.~Voloshin, Phys.Rev. {\bf D86} (2012) 093016,
  [\href{http://xxx.lanl.gov/abs/1208.4303}{{\tt arXiv:1208.4303}}].

\bibitem{Korchin:2013ifa}
A.~Y. Korchin and V.~A. Kovalchuk, Phys.Rev.
  {\bf D88} (2013), no.~3 036009,
  [\href{http://xxx.lanl.gov/abs/1303.0365}{{\tt arXiv:1303.0365}}].

\bibitem{Bishara:2013vya}
F.~Bishara, Y.~Grossman, R.~Harnik, D.~J. Robinson, J.~Shu, {\em et.~al.}, JHEP {\bf 1404} (2014) 084, [\href{http://xxx.lanl.gov/abs/1312.2955}{{\tt
  arXiv:1312.2955}}].


\bibitem{Chen:2014ona}
  Y.~Chen, A.~Falkowski, I.~Low and R.~Vega-Morales,
  Phys.\ Rev.\ D {\bf 90} (2014) 11,  113006, \href{http://xxx.lanl.gov/abs/1405.6723}{{\tt arXiv:1405.6723}}.

\bibitem{Freitas:2012kw}
A.~Freitas and P.~Schwaller, Phys.Rev. {\bf D87} (2013), no.~5 055014,
  [\href{http://xxx.lanl.gov/abs/1211.1980}{{\tt arXiv:1211.1980}}].


\bibitem{Djouadi:2013qya}
A.~Djouadi and G.~Moreau, {\it {The couplings of the Higgs boson and its CP
  properties from fits of the signal strengths and their ratios at the 7+8 TeV
  LHC}},  Eur.Phys.J. {\bf C73} (2013) 2512,
  [\href{http://xxx.lanl.gov/abs/1303.6591}{{\tt arXiv:1303.6591}}].


\bibitem{Belusca-Maito:2014dpa}
H.~Belusca-Maito, \href{http://xxx.lanl.gov/abs/1404.5343}{{\tt
  arXiv:1404.5343}}.
  
\bibitem{Shu:2013uua}
J.~Shu and Y.~Zhang, Phys.\ Rev.\ Lett.\ {\bf 111} (2013) 9, 091801,
\href{http://xxx.lanl.gov/abs/1304.0773}{{\tt arXiv:1304.0773}}.

\bibitem{Cheung:2013kla}
K.~Cheung, J.~S.~Lee and P.~-Y.~Tseng, JHEP {\bf 1305} (2013) 134
\href{http://xxx.lanl.gov/abs/1302.3794}{{\tt  arXiv:1302.3794}}.

\bibitem{Dwivedi:2015nta}
  S.~Dwivedi, D.~K.~Ghosh, B.~Mukhopadhyaya and A.~Shivaji,
  arXiv:1505.05844 [hep-ph].


\bibitem{LRSM1}
J.C.Pati, A.Salam, Phys. Rev. {\bf D10}, 275(1974).

\bibitem{LRSM2}
R.N.Mohapatra, J.C.Pati, Phys. Rev. {\bf D11}, 566(1975);
R.N.Mohapatra, J.C.Pati, Phys. Rev. {\bf D11}, 2558(1975);
G.Senjanovic, R.N.Mohapatra, Phys.Rev.{\bf D12}, 1502(1975).

\bibitem{Georgi:1984af}
H.~Georgi and D.~B. Kaplan,  Phys.Lett. {\bf B145} (1984) 216.


\bibitem{Yepes:2015zoa}
  J.~Yepes, arXiv:1507.03974 [hep-ph].
  

\bibitem{Zhang:2007xy}
  Y.~Zhang, S.~Z.~Wang, F.~J.~Ge and Q.~Wang,
  Phys.\ Lett.\ B {\bf 653} (2007) 259
  [arXiv:0704.2172 [hep-ph]]


\bibitem{Wang:2008nk}
  S.~Z.~Wang, S.~Z.~Jiang, F.~J.~Ge and Q.~Wang,
  JHEP {\bf 0806} (2008) 107
  [arXiv:0805.0643 [hep-ph]].


\bibitem{Appelquist:1980vg}
T.~Appelquist and C.~W. Bernard,  Phys. Rev. {\bf D22} (1980) 200.

\bibitem{Longhitano:1980iz}
A.~C. Longhitano,  Phys. Rev. {\bf D22} (1980) 1166.

\bibitem{Longhitano:1980tm}
A.~C. Longhitano,  Nucl. Phys. {\bf B188} (1981) 118.

\bibitem{Feruglio:1992wf}
F.~Feruglio,  Int.J.Mod.Phys. {\bf A8} (1993) 4937--4972,
  [\href{http://xxx.lanl.gov/abs/hep-ph/9301281}{{\tt hep-ph/9301281}}].

\bibitem{Appelquist:1993ka}
T.~Appelquist and G.-H. Wu,  Phys.Rev. {\bf D48} (1993) 3235--3241,
  [\href{http://xxx.lanl.gov/abs/hep-ph/9304240}{{\tt hep-ph/9304240}}].


\bibitem{Gavela:2014vra}
  M.~B.~Gavela, J.~Gonzalez-Fraile, M.~C.~Gonzalez-Garcia, L.~Merlo, S.~Rigolin and J.~Yepes, JHEP {\bf 1410} (2014) 44
  [arXiv:1406.6367 [hep-ph]].


\bibitem{Alonso:2012px}
R.~Alonso, M.~B.~Gavela, L.~Merlo, S.~Rigolin, and J.~Yepes, 
Phys.Lett. {\bf B722} (2013) 330--335,
  [\href{http://xxx.lanl.gov/abs/1212.3305}{{\tt arXiv:1212.3305}}].


\bibitem{Agashe:2006at}
  K.~Agashe, R.~Contino, L.~Da Rold {\it et al.},
  %``A Custodial symmetry for Zb anti-b,''
  Phys.\ Lett.\  {\bf B641} (2006) 62-66.
  [arXiv:0605341 [hep-ph]].


\bibitem{Brivio:2013pma}
I.~Brivio, T.~Corbett, O.~Eboli, M.~B.~Gavela, J.~Gonzalez-Fraile, {\em et.~al.}, JHEP {\bf 1403} (2014) 024,
  [\href{http://xxx.lanl.gov/abs/1311.1823}{{\tt arXiv:1311.1823}}].


\bibitem{Alonso:2012pz}
R.~Alonso, M.~B.~Gavela, L.~Merlo, S.~Rigolin, and J.~Yepes,  Phys.Rev. {\bf D87}  (2013) 055019, [\href{http://xxx.lanl.gov/abs/1212.3307}{{\tt  arXiv:1212.3307}}].


\bibitem{Buchalla:2012qq}
G.~Buchalla and O.~Cat\`a, JHEP {\bf 1207} (2012) 101
 [\href{http://xxx.lanl.gov/abs/1203.6510}{{\tt arXiv:1203.6510}}].


\bibitem{Buchalla:2013rka}
  G.~Buchalla, O.~Cata and C.~Krause,
  Nucl.\ Phys.\ B {\bf 880} (2014) 552
  [arXiv:1307.5017 [hep-ph]].


\bibitem{Buchalla:2013eza}
  G.~Buchalla, O.~Cata and C.~Krause,
  Phys.\ Lett.\ B {\bf 731} (2014) 80
  [arXiv:1312.5624 [hep-ph]].


\bibitem{Brivio:2015hua}
  I.~Brivio,
  arXiv:1505.00637 [hep-ph].

\bibitem{Cvetic:1988ey}
G.~Cvetic and R.~Kogerler,  
Nucl.Phys. {\bf B328} (1989)  342.


\bibitem{Alonso:2012jc}
R.~Alonso, M.~Gavela, L.~Merlo, S.~Rigolin, and J.~Yepes, JHEP {\bf 1206} (2012) 076,
  [\href{http://xxx.lanl.gov/abs/1201.1511}{{\tt arXiv:1201.1511}}].


\bibitem{Brivio:2014pfa}
  I.~Brivio, O.~J.~P.~Eboli, M.~B.~Gavela, M.~C.~Gonzalez-Garcia, L.~Merlo and S.~Rigolin, JHEP {\bf 1412} (2014) 004, \href{http://xxx.lanl.gov/abs/1405.5412}{{\tt arXiv:1405.5412}}.


\bibitem{Contino:2011np}
R.~Contino, D.~Marzocca, D.~Pappadopulo, and R.~Rattazzi, 
JHEP {\bf 1110} (2011) 081, [\href{http://xxx.lanl.gov/abs/1109.1570}{{\tt arXiv:1109.1570}}].


\bibitem{Alonso:2014wta}
  R.~Alonso, I.~Brivio, B.~Gavela, L.~Merlo and S.~Rigolin,
  JHEP {\bf 1412} (2014) 034,
  \href{http://xxx.lanl.gov/abs/1409.1589}{{\tt arXiv:1409.1589}}.


\bibitem{Shu:2015cxm}
  J.~Shu and J.~Yepes, [\href{http://xxx.lanl.gov/abs/1512.09310}{{\tt arXiv:1512.09310}}].


\bibitem{Hagiwara:1986vm}
K.~Hagiwara, R.~Peccei, D.~Zeppenfeld, and K.~Hikasa, {\it {Probing the Weak
  Boson Sector in E+ E- --> W+ W-}},  Nucl.Phys. {\bf B282} (1987) 253.

\bibitem{Appelquist:2004mn}
T.~Appelquist, M.~Piai and R.~Shrock,
{\it {Lepton Dipole Moments in Extended Technicolor Models}},
Phys.\ Lett.\ B {\bf 593} (2004) 175,
[\href{http://xxx.lanl.gov/abs/hep-ph/0401114}{{\tt arXiv:hep-ph/0401114}}].

\bibitem{Marciano:1986eh}
W.~J. Marciano and A.~Queijeiro, {\it {Bound on the W Boson Electric Dipole
  Moment}},  Phys.Rev. {\bf D33} (1986) 3449.

\bibitem{Baron:2013eja}
{\bf ACME} Collaboration, J.~Baron {\em et.~al.}, {\it {Order of Magnitude
  Smaller Limit on the Electric Dipole Moment of the Electron}},  Science {\bf
  343} (2014), no.~6168 269--272,
  [\href{http://xxx.lanl.gov/abs/1310.7534}{{\tt arXiv:1310.7534}}].

\bibitem{Baker:2006ts}
C.~Baker, D.~Doyle, P.~Geltenbort, K.~Green, M.~van~der Grinten, {\em et.~al.},
  {\it {An Improved Experimental Limit on the Electric Dipole Moment of the
  Neutron}},  Phys.Rev.Lett. {\bf 97} (2006) 131801,
  [\href{http://xxx.lanl.gov/abs/hep-ex/0602020}{{\tt hep-ex/0602020}}].

\bibitem{Dawson:2013owa}
S.~Dawson, S.~K. Gupta, and G.~Valencia, {\it {CP violating anomalous couplings
  in $W\gamma$ and $Z\gamma$ production at the LHC}},  Phys.Rev. {\bf D88}
  (2013), no.~3 035008, [\href{http://xxx.lanl.gov/abs/1304.3514}{{\tt
  arXiv:1304.3514}}].

\bibitem{Abbiendi:2000ei}
{\bf OPAL} Collaboration, G.~Abbiendi {\em et.~al.}, {\it {Measurement of $W$
  boson polarizations and CP violating triple gauge couplings from $W^{+}
  W^{-}$ production at LEP}},  Eur.Phys.J. {\bf C19} (2001) 229--240,
  [\href{http://xxx.lanl.gov/abs/hep-ex/0009021}{{\tt hep-ex/0009021}}].

\bibitem{Abdallah:2008sf}
{\bf DELPHI} Collaboration, J.~Abdallah {\em et.~al.}, {\it {Study of W boson
  polarisations and Triple Gauge boson Couplings in the reaction e+e- --$>$W+W-
  at LEP 2}},  Eur.Phys.J. {\bf C54} (2008) 345--364,
  [\href{http://xxx.lanl.gov/abs/0801.1235}{{\tt arXiv:0801.1235}}].

\bibitem{Schael:2004tq}
{\bf ALEPH} Collaboration, S.~Schael {\em et.~al.}, {\it {Improved measurement
  of the triple gauge-boson couplings gamma W W and Z W W in e+ e-
  collisions}},  Phys.Lett. {\bf B614} (2005) 7--26.

\bibitem{Chatrchyan:2013mxa}
{\bf CMS} Collaboration, S.~Chatrchyan {\em et.~al.}, {\it {Measurement of the
  properties of a Higgs boson in the four-lepton final state}},  Phys.Rev. {\bf
  D89} (2014) 092007, [\href{http://xxx.lanl.gov/abs/1312.5353}{{\tt
  arXiv:1312.5353}}].

\bibitem{Shu:2016exh}
  J.~Shu and J.~Yepes, [\href{http://xxx.lanl.gov/abs/1601.06891}{{\tt arXiv:1601.06891}}]

\end{thebibliography}
%\bibliographystyle{BiblioStyleLetter}

\providecommand{\href}[2]{#2}\begingroup\raggedright\endgroup

\end{document}